\documentclass[aps,
prx,
twocolumn,
superscriptaddress,
10pt,
]{revtex4-2}

\usepackage[T1]{fontenc}
\usepackage{lmodern}
\usepackage{color}
\usepackage{amsmath}
\usepackage{amssymb}
\usepackage{graphicx}

\usepackage[colorlinks=true,allcolors=blue]{hyperref}
\usepackage{mathtools}
\usepackage{makecell,tabularx}
\usepackage[normalem]{ulem}
\usepackage{natbib}
\usepackage{textgreek}
 
\usepackage{times}
\usepackage{physics}
\usepackage{xcolor}
\usepackage{braket}
\usepackage{upgreek}
\usepackage{setspace}
\usepackage{float}
\makeatother

\begin{document}

\title{Protomon: A Multimode Qubit in the Fluxonium Molecule}

\author{Shashwat Kumar}
\email{sk8995@princeton.edu}
\thanks{These authors contributed equally to this work}
\affiliation{Department of Electrical and Computer Engineering, Princeton University, Princeton, New Jersey 08544, USA}

\author{Xinyuan You}
\altaffiliation[Present address: ]{Superconducting Quantum Materials and Systems Center, Fermi National Accelerator Laboratory (FNAL), Batavia, IL 60510, USA}
\thanks{These authors contributed equally to this work}
\affiliation{Northwestern--Fermilab Center for Applied Physics and Superconducting Technologies, Northwestern University, Evanston, Illinois 60208, USA}
\affiliation{Graduate Program in Applied Physics, Northwestern University, Evanston, Illinois 60208, USA}

\author{Xanthe Croot}
\thanks{These authors contributed equally to this work}
\affiliation{Department of Physics, Princeton University, Princeton, New Jersey 08544, USA}

\author{Tianpu Zhao}
\affiliation{Graduate Program in Applied Physics, Northwestern University, Evanston, Illinois 60208, USA} 

\author{Danyang Chen}
\affiliation{Department of Physics and Astronomy, Northwestern University, Evanston, Illinois 60208, USA}

\author{Sara Sussman}
\affiliation{Department of Physics, Princeton University, Princeton, New Jersey 08544, USA}

\author{Anjali Premkumar}
\affiliation{Department of Electrical and Computer Engineering, Princeton University, Princeton, New Jersey 08544, USA}

\author{Jacob Bryon}
\affiliation{Department of Electrical and Computer Engineering, Princeton University, Princeton, New Jersey 08544, USA}

\author{Jens Koch}
\affiliation{Department of Physics and Astronomy, Northwestern University, Evanston, Illinois 60208, USA}

\author{Andrew A. Houck}
\email{aahouck@princeton.edu}
\affiliation{Department of Electrical and Computer Engineering, Princeton University, Princeton, New Jersey 08544, USA}

\begin{abstract}
Qubits that are intrinsically insensitive to depolarization and dephasing errors promise to significantly reduce the overhead of fault-tolerant quantum computing. 
At their optimal operating points, the logical states of these qubits exhibit both exponentially suppressed matrix elements and sweet spots in energy dispersion, rendering the qubits immune to depolarization and dephasing, respectively.
We introduce a multimode qubit, the \textit{protomon}, encoded in a fluxonium molecule circuit. 
Compared to the closely related $0$--$\ensuremath{\pi}$ qubit, the protomon offers several advantages in theory: resilience to circuit parameter disorder, minimal dephasing from intrinsic harmonic modes, and no dependence on static offset charge. 
As a proof of concept, we realize four protomon qubits. 
By tuning the qubits to various operating points identified with calibrated two-tone spectroscopy, we measure depolarization times ranging from 64 to 73 μs and dephasing times between 0.2 to 0.5 μs for one selected qubit. The discrepancy between the relatively short measured coherence times and theoretical predictions is not fully understood.
This calls for future studies investigating the limiting noise factors, informing the direction for improving coherence times of the protomon qubit.

\end{abstract}

\maketitle

\section{Introduction}

To execute large-scale algorithms on a quantum computer, it is essential to correct errors that occur during the computation. Quantum error correction protocols make this possible, provided that qubit error rates are below a protocol-dependent threshold. A primary goal, therefore, for any quantum computing platform is to suppress qubit error rates below this fault-tolerant threshold. 
Although many physical qubit platforms have achieved this error threshold~\cite{google2023suppressing,PhysRevLett.127.130505,PhysRevX.11.041058,zajac2021spectator,evered2023high}, further improvement in coherence time is necessary to significantly reduce the overhead in fault-tolerant quantum computing, specifically reducing the number of physical qubits required to encode a logical qubit~\cite{Knill2005}.

Efforts to improve coherence times in superconducting qubits have adopted a two-pronged approach. Advances in materials science and qubit fabrication mitigate noise in the immediate qubit environment~\cite{Wang2015, Place2021,Wang2022,bal_systematic_2024}, while carefully engineering coupling between the qubit and environment suppress qubit sensitivity to noise. The latter strategy has motivated recent experiments to realize qubits in both conventional platforms, such as superconducting circuits~\cite{kalashnikov2020bifluxon,Gyenis2021, GyenisReview2021}, as well as more exotic topological implementations~\cite{Kitaev2003} using hybrid semiconductor-superconducting architectures~\cite{PhysRevLett.125.056801,Schrade2022}. 
Such qubits are resistant to both depolarization and dephasing errors. These error rates are characterized by the inverse of the qubit depolarization time $T_1$ and the inverse of the qubit dephasing time $T_\phi$, respectively.
Since the qubit depolarization time is inversely proportional to the matrix elements of the local qubit operators coupled to the bath, a longer $T_1$ can be achieved if logical states possess disjoint wavefunctions or have identical parities~\cite{GyenisReview2021}.
Similarly, the dephasing time is determined by how sensitive the qubit's energy splitting is to noise parameters. By operating at sweet spots where the qubit energy spectrum exhibits a (first-order) vanishing derivative with respect to the noise parameter, the dephasing time $T_\phi$ can be greatly enhanced. 

In general, single-mode superconducting qubits like the transmon and fluxonium provide protection against either depolarization or dephasing processes. 
However, achieving both simultaneously is challenging~\cite{GyenisReview2021}. For example, a fluxonium operating at half flux quantum exhibits first-order insensitivity to flux noise but it lacks protection against depolarization noise.
In this work, we present the \textit{protomon}, a superconducting qubit encoded in a multimode fluxonium molecule circuit~\cite{kou2017}.
This concept was proposed in Ref.~\onlinecite{YouThesis}, and has since been refined through the incorporation of Floquet engineering techniques~\cite{thibodeau2024floquet}.
With carefully designed Hamiltonian parameters, there are three sweet spots that exhibit both disjoint support and vanishing first-order energy dispersion with respect to external flux.
Compared to the closely related $0$--$\ensuremath{\pi}$ qubit\cite{PhysRevA.87.052306,Groszkowski2018,Gyenis2021}, the protomon offers several advantages in theory. First, the qubit's coherence properties are insensitive to disorder in circuit parameters. Second, the high energy of the spurious harmonic mode compared to the thermal energy suppresses photon shot noise. Moreover, replacing the shunt capacitor in the $0$--$\ensuremath{\pi}$ qubit with an inductor eliminates offset charge dependence.

We structure this paper as follows: In Sec.~\ref{section:theory}, we introduce the fluxonium molecule circuit and discuss its resilience against noise, supported by numerical calculations of the coherence times.
In Sec.~\ref{section:experiment}, we detail the experimental realization of the protomon, present spectroscopic data to demonstrate flux dependencies of the level spectrum, and characterize qubit coherences. 
We share our conclusions and outlook in Sec.~\ref{sec:conclusion}. Supplementary details are included in the appendices.

\section{Theory of the protomon qubit}
\label{section:theory}
\begin{figure}[t]
\includegraphics[width=\columnwidth]{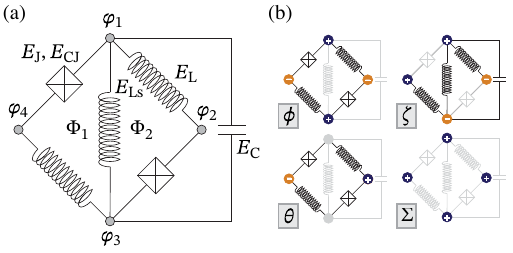}
\protect\caption{\label{fig:figure1}
(a) Circuit schematic of the protomon. Similar to the fluxonium molecule, the circuit consists of two fluxonia, each comprising a Josephson junction (energy $E_{\text J}$) and an inductor (energy $E_{\text L}$). These fluxonia share a common inductor (energy $E_{\text{Ls}}$). Each fluxonium forms a unique loop threaded by fluxes $\Phi_{1,2}$. 
(b) Circuit schematic of the normal modes of the linearized circuit. The sign of the normal-mode amplitude is indicated at each node. 
}
\end{figure}

The protomon qubit is encoded in a subspace of the fluxonium molecule~\cite{kou2017}, utilizing a carefully engineered parameter regime. 
The circuit diagram, shown in Fig.~\ref{fig:figure1}(a), consists of four nodes and two independent loops. Each loop ($i$ = 1, 2) includes a Josephson junction with capacitance $C_{\text{J},i}$ and junction energy $E_{\text{J},i}$, along with a superinductor denoted by $E_{\text{L},i}$. 
A shunt superinductor $E_\text{Ls}$ couples the two loops, and allows for individual control of the magnetic flux $\Phi_{i}$ through each loop. 
A small but finite capacitance $C$ is introduced between nodes 1 and 3 to account for the residual capacitance, thus avoiding singular Lagrangian~\cite{PhysRevX.13.021017}. 
For simplicity, we assume the parameters for the Josephson junctions and superinductors in both loops are identical. The presence of disorder in circuit parameters does not lead to any qualitative changes and is discussed in detail in the Appendix~\ref{app:disorder}.

\begin{figure*}[t]
\includegraphics[width=\textwidth]{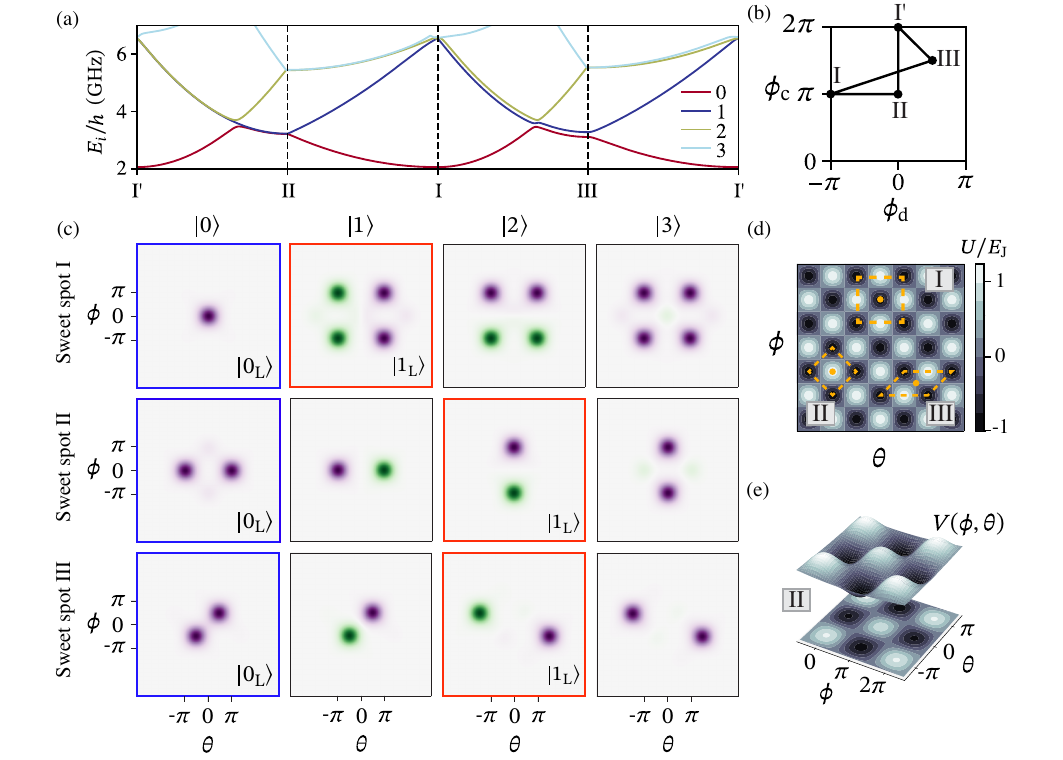}
\protect\caption{\label{fig:figure2}
Spectra and wavefunctions of the protomon qubit. 
(a) The lowest four eigenenergies along the trajectory shown in (b). 
(b) Inversion-symmetric points (sweet spots) of the Hamiltonian in the plane of common ($\phi_\text{c}$) and differential ($\phi_\text{d}$) external fluxes. The sweet spots are labeled I [$(\phi_\text{c}, \phi_\text{d}) = (\pi, -\pi)$], I' [$(2\pi, 0)$], II [$(\pi, 0)$], and III [$(3\pi/2, -\pi/2)$]. The Hamiltonians at sweet spots I and I' are identical.
(c) Wavefunctions of the lowest four energy eigenstates of the reduced Hamiltonian [Eq.~(\ref{eq:fm:heff})] at sweet spots I, II and III, represented in the $\phi$--$\theta$ basis. The logical 0 (1) states are highlighted with blue (red) boxes.
(d) Periodic potential energy landscape solely from the Josephson junctions i.e., $- 2 E_\text{J}\cos(\hat{\phi}+\phi_\text{c})\cos(\hat{\theta}+\phi_\text{d})$.
The dashed orange lines connect local potential minima where the eigenstates at the respective sweet spot are localized. The filled dots denote the minima of the potential from the inductor alone.
(e) Potential energy landscape and its two-dimensional projection, when the circuit is operated at the double sweet spot II. 
[Circuit parameters: $E_\text{J}$/$h$ = 11 GHz, $E_\text{ CJ}$/$h$ = 2.5 GHz, $E_\text{C}$/$h$ = 50 GHz, and $E_{\text L}/h$ = $E_{\text{Ls}}$/$h$ = 0.36 GHz]
}

\end{figure*}

\subsection{Full and reduced circuit Hamiltonian}
To quantize the circuit, we first construct the Lagrangian of the circuit in terms of the node variables $\varphi_{1,2,3,4}$, as depicted in Fig.~\ref{fig:figure1}(a). 
To simplify the Lagrangian, we implement a variable transformation and introduce a set of variables $\phi, \theta, \zeta, \Sigma$, with the normal-mode amplitudes illustrated in Fig.~\ref{fig:figure1}(b). 
Notably, the $\Sigma$ mode does not induce a voltage drop across any circuit element, rendering it cyclic and irrelevant to the circuit dynamics. Details of the Lagrangian formulation and variable transformation are provided in Appendix~\ref{app:full_ham}.
Applying the Legendre transformation followed by canonical quantization yields the circuit Hamiltonian
\begin{equation}\label{eq:fm:ham}
    \begin{split}
    \hat{\mathcal{H}} = \,& 2 E_\text{CJ} (\hat{n}_\phi^2 + \hat{n}_\theta^2) + 4 E_\text{C}\hat{n}_\zeta^2 + E_\text{L} \left[\hat{\phi}^2 + (\hat{\theta}-\hat{\zeta})^2 \right]\\
    & + \dfrac{1}{2}E_\text{Ls}\hat{\zeta}^2 -2E_\text{J}\cos(\hat{\phi}+\phi_\text{c})\cos(\hat{\theta}+\phi_\text{d}),
    \end{split}
\end{equation}
where $\hat{n}_\phi$, $\hat{n}_\theta$, and $\hat{n}_\zeta$ denote the operators conjugate to $\hat{\phi}$, $\hat{\theta}$, and $\hat{\zeta}$, respectively. The charging energies are defined as $E_\text{CJ}=e^2/2C_\text{J}$ and $E_\text{C}=e^2/2C$. Additionally, the reduced common and differential external fluxes threading the two loops are denoted by $\phi_\text{c} = (\Phi_\text{1} + \Phi_\text{2})/(2\phi_0)$ and $\phi_\text{d} = (\Phi_\text{1} - \Phi_\text{2})/(2\phi_0)$, respectively. The reduced flux quantum is $\phi_0=\hbar/2e$. 

It is instructive to compare the above Hamiltonian with the one for the $0$--$\ensuremath{\pi}$ qubit~\cite{Gyenis2021}. 
First, the incorporation of an additional shunt inductor leads to only one island within the circuit configuration. As a result, all three variables (\text{i.e.}, $\theta$, $\phi$ and $\zeta$) are extended variables, in contrast to the case of the $0$--$\ensuremath{\pi}$ qubit, where $\theta$ is a periodic variable. 
Therefore, we can perform a gauge transformation to eliminate the dependence on static offset charges in the Hamiltonian and convert the spectrum of the relevant noise from $1/f$ to Ohmic, thereby protecting the qubit from charge noise~\cite{Koch2009}.
Furthermore, the $\zeta$ mode couples to the $\theta$ mode (and indirectly to $\phi$ mode) without the presence of disorder. This is different from the $0$--$\ensuremath{\pi}$ qubit where the $\zeta$ mode is fully decoupled when all circuit elements are identical~\cite{Groszkowski2018}. However, as we show in the following, an effective Hamiltonian describing only the $\theta$ and $\phi$ degrees of freedom is sufficient in the intended parameter regime. 

From the full Hamiltonian in Eq.~\eqref{eq:fm:ham}, we extract the resonant frequency of the \(\zeta\) mode to be \(\hbar\omega_\zeta = \sqrt{8E_{\text{C}}(2E_{\text{L}} + E_{\text{Ls}})}\). The small capacitance \(C\) elevates \(\omega_\zeta\) to the largest energy scale relevant for the qubit dynamics. 
Consequently, the Hilbert space partitions into distinct manifolds corresponding to excitations of the harmonic \(\zeta\) mode. 
At low temperatures, excitation of the harmonic mode is substantially suppressed. 
Thus, we can assume the system is always in the ground state of the $\zeta$ mode. 
To approximate the low-energy dynamics of the full Hamiltonian, we apply the Schrieffer--Wolf transformation (see Appendix~\ref{app:reduced_ham} for details), resulting in a reduced Hamiltonian for the \(\theta\) and \(\phi\) degrees of freedom:
\begin{equation}\label{eq:fm:heff}
    \begin{split}
        \hat{\mathcal{H}}_\text{eff} =&  2 E_\text{CJ} \hat{n}_\theta^2  +2 E_\text{CJ} \hat{n}_\phi^2 + \dfrac{E_\text{L}E_\text{Ls}}{2E_\text{L} + E_\text{Ls} } \hat{\theta}^2  + E_\text{L} \hat{\phi} ^2 \\
        & - 2 E_\text{J}\cos(\hat{\phi}+\phi_\text{c})\cos(\hat{\theta}+\phi_\text{d}).
    \end{split}
\end{equation}
This describes a particle in a two-dimensional potential, which consists of a double cosine potential and an anisotropic parabolic potential. 
The relative position of the center of the paraboloid with respect to the cosine potential is determined by the applied fluxes $\phi_\text{c}$ and $\phi_\text{d}$. 
The parameter regime discussed throughout the paper follows $E_\text{L} \ll E_\text{J}$ and $E_\text{CJ}< E_\text{J}$. The first condition ensures the existence of multiple wells, while the latter suppresses the tunneling amplitude between the potential wells.

\subsection{Coherence properties of the protomon qubit}
To investigate the coherence properties of the protomon qubit, it is essential to analyze both the energy spectrum and the wavefunctions derived from the effective Hamiltonian in Eq.~\eqref{eq:fm:heff}. We focus on the lowest four eigenstates of this reduced Hamiltonian, denoted as $|0\rangle$, $|1\rangle$, $|2\rangle$, and $|3\rangle$. 
After numerically diagonalizing the Hamiltonian with \texttt{scQubits}~\cite{Groszkowski2021, Chitta2022}, we show in Fig.~\ref{fig:figure2}(a) the four lowest eigenenergies as a function of the applied common flux \(\phi_{\text{c}}\) and differential flux \(\phi_{\text{d}}\). The horizontal axis represents a specific trajectory within the $\phi_{\text{c}}$--$\phi_{\text{d}}$ plane, connecting the points I, II, and III [see Fig.~\ref{fig:figure2}(b)]. The Hamiltonian exhibits inversion symmetry at those points, as it is invariant under the transformations \(\phi_{1,2} \rightarrow -\phi_{1,2}\) and \(\phi_{1,2} \rightarrow \phi_{1,2} + 2\pi\). 
At these inversion-symmetric points, the energy differences between eigenstates reach extremal values, indicating vanishing first-order derivatives to the applied fluxes. We refer to these symmetry points as \textit{double} flux sweet spots. Additionally, the positions of avoided crossings are marked with ellipses, with the size of each ellipse proportional to the magnitude of the energy gap at the crossing.
The existence of three double flux sweet spots is promising for constructing a qubit that is first-order insensitive to the flux noise, which is typically the dominant source of pure dephasing for flux-tunable qubits.

In order to protect against depolarization noise, the wavefunctions of the logical states must exhibit disjoint support. Fig.~\ref{fig:figure2}(c) displays the wavefunctions of the lowest four eigenstates in the $\phi$--$\theta$ basis, when operating at sweet spots I, II, and III, respectively. 
To help visualize the structure of the wavefunctions, we show in Fig.~\ref{fig:figure2}(d) an overlay of the double-cosine potential. 
The logical states $|0_\text{L}\rangle$ and $|1_\text{L}\rangle$ at each operating point are highlighted with red and blue boxes, respectively. Note that they are not necessarily the lowest two eigenstates.
As an example, we examine the wavefunctions at sweet spot II, where the corresponding potential energy landscape is illustrated in Fig.~\ref{fig:figure2}(e). The ground and first excited states, $|0\rangle$ and $|1\rangle$, consist of symmetric and anti-symmetric superpositions of states localized in two potential wells along the $\theta$ direction. Likewise, the second and third excited states, $|2\rangle$ and $|3\rangle$, are superpositions of states occupying wells along the $\phi$ direction. 
This wavefunction structure can be intuitively understood through a simple tunneling model (see Appendix~\ref{app:hopping}). 
This model also explains the structure in the spectrum [Fig.~\ref{fig:figure2}(a)], where avoided crossings occur between $|0\rangle$ and $|1\rangle$ and between $|2\rangle$ and $|3\rangle$. 
In this configuration, with the logical states defined as $|0\rangle$ and $|2\rangle$, their wavefunctions exhibit disjoint support, as they occupy distinct wells in the two-dimensional potential landscape. Consequently, the matrix elements of any local operator, which typically couples to noise channels such as capacitive loss, inductive loss, and quasiparticle loss, are exponentially suppressed between these logical states. This ensures a significant reduction in depolarization rates within the logical subspace due to such noise processes.
However, there exist leakages outside of the logical subspace, i.e., $|0\rangle\to|1\rangle$ and $|2\rangle\to|3\rangle$. The wavefunctions of the states involved in those leakages have strong overlap, limiting the lifetime of the protomon qubit, if no post-selection technique is applied.

\begin{figure}[t]
\includegraphics[width=1\columnwidth]{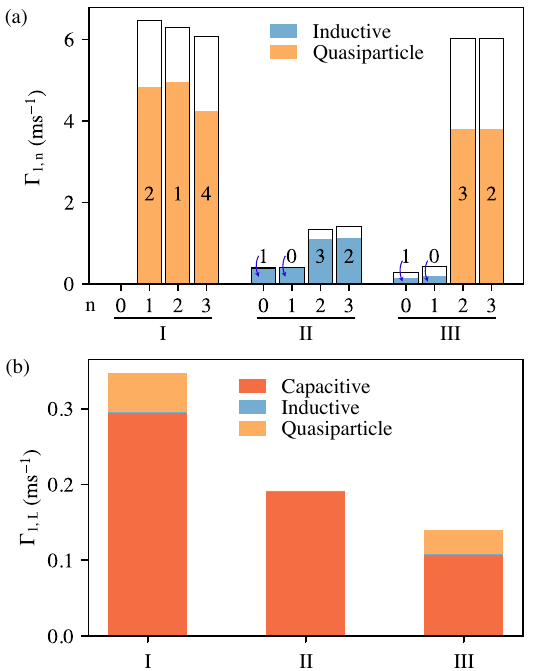}
\protect\caption{\label{fig:coherence} 
Coherence analysis of the protomon circuit at sweet spots I, II, and III.
(a) The black-outlined bars represent $\Gamma_{1,n}$, the sum of rates associated with transitions originating from each state $\ket{n}$ (labeled on the horizontal axis). 
The colored portion of the bars indicates the dominant contribution of the transition rates among all the noise channels and all the destination states; the color encodes the noise channel and the corresponding destination is labeled on each bar. 
(b) The qubit logical lifetime \(\Gamma_{1,\text{L}} \) between the logical states $\ket{0_\text{L}}$ and $\ket{1_\text{L}}$ chosen at different sweet spots, with vertical bars representing contributions from the various noise channels.
} 
\end{figure}

Next, we numerically calculate the qubit coherence times. 
Figure~\ref{fig:coherence}(a) presents the golden-rule type decay rates for each state (i.e., $|0\rangle$, $|1\rangle$, $|2\rangle$, $|3\rangle$) at various sweet spots (i.e., I, II, and III). 
Specifically, the decay rate for transitions originating from a state \(|n\rangle\) (indicated on the horizontal axis) is given by the sum of rates for transitions away from it, i.e., \(\Gamma_{1,n} = \sum_{j \neq n}\Gamma_{1,nj}^\text{cap.} + \Gamma_{1,nj}^\text{ind.} + \Gamma_{1,nj}^\text{qp}\), where each term $\Gamma_{1,nj}^k$ denotes the transition to state \(|j\rangle\) induced by the noise channel $k$. 
For simplicity, only the dominant contribution is shown explicitly, with the color indicating the noise source and the label on each bar representing the state $|j\rangle$. The subdominant contributions are grouped and shown in gray. 
For example, at sweet spot II, the decay rates for the lowest four states are in the millisecond range, dominated by rates for transitions between adjacent levels, i.e., \(|0\rangle \leftrightarrow |1\rangle\) and \(|2\rangle \leftrightarrow |3\rangle\) [see level structure in Fig.~\ref{fig:figure2}(a)]. 
These transitions are dominated by inductive loss, as the contribution from capacitive loss is suppressed due to the vanishing charge matrix element. 
A similar approach can be applied to calculate the pure dephasing times, which are generally limited by flux noise. However, we refrain from providing quantitative predictions here, as the treatment of second-order flux noise at the sweet spots remains a subject of ongoing discussion~\cite{Groszkowski2018}.

\subsection{Coherence of logical subspace with possible erasure check}

The analysis above demonstrates that the protomon qubit can achieve millisecond-scale lifetimes at the sweet spots. The dominant noise channels limiting the lifetime are attributed to leakage into adjacent levels. By actively detecting such leakage, it is possible to convert these errors into erasure errors, which can be efficiently corrected~\cite{Kubica2023,Chou2023,Teoh2023,Levine2024}. 
In Fig.~\ref{fig:coherence}(b), we estimate the coherence of the logical subspace, assuming that leakage has been corrected. For example, at sweet spot II, where \(|0\rangle\) and \(|2\rangle\) are chosen as the logical states, the qubit logical lifetime (\(\Gamma_{1,\text{L}} = \Gamma_{1,02} + \Gamma_{1,20}\)) is calculated to exceed tens of milliseconds. This is a direct consequence of the disjoint support of the wavefunctions. The primary noise contributions are due to capacitive loss, as indicated by the color of the vertical bar.

\section{Experimental Realization}
\label{section:experiment}

\begin{figure}[t]
\includegraphics[width=1\columnwidth]{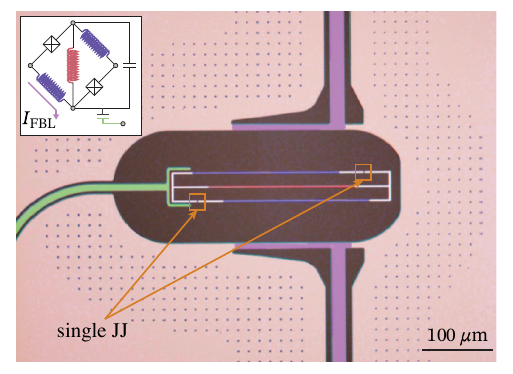}
\protect\caption{\label{fig:figure3}Optical microscope image of Device 1. The protomon, fabricated from Dolan junctions, is capacitively coupled to a coplanar waveguide resonator for readout. The device is depicted in false colors for clarity, corresponding to the colors used in the circuit diagram (inset). One flux bias line and a global magnet are utilized for controlling the external flux.
} 
\end{figure}

We realize the protomon qubit in four unique circuits (Devices 1 to 4), with circuit parameters in the target regime.  All devices are fabricated on sapphire substrates and surrounding circuitry, including capacitively coupled co-planar waveguide readout resonators, is defined in thin-film tantalum~\cite{Place2021} using photolithography, as shown in Fig.~\ref{fig:figure3}. While Devices 1 to 3 are all-aluminum qubits, Device 4 integrates a small area of tantalum into the capacitive component of the circuit. In all devices, dense arrays of Josephson junctions achieve the large inductances required for the target $E_{\text L}$~\cite{Masluk2012, Manucharyan2009}, which must be significantly lower than the Josephson energies of individual junctions to ensure tunneling is suppressed between adjacent wells. All Josephson junctions are fabricated via the Dolan bridge technique~\cite{Dolan1977}. Junctions are connected by large metallic islands corresponding to nodes of the qubit circuit. Each qubit is capacitively coupled to a nearby coplanar waveguide resonator, allowing measurement of the qubit states via dispersive readout.

\begin{table*}[t]
  \centering
  \caption{Device parameters and coherence times at sweet spots I and II (labeled in square brackets). 
  The definitions of measured decoherence times $T_{1}^{\text{s}}$ and $T_{2\text{R}}^{\text{s}}$ are detailed in Sec.~\ref{section:experiment} and Appendix~\ref{sec: noise-subspace}.
  }
  \label{table:devices_coherences}
  \begin{ruledtabular}
    \begin{tabular}{cccccccccc}
      & $E_\text{J}/h$ (GHz) & $E_\text{L}/h$ (GHz) & $E_\text{Ls}/h$ (GHz) & ${E_\text{CJ}/h}$ (GHz) & ${E_\text{C}/h}$ (GHz) & $T_{1}^{\text{s}}[\text{I}]$ (μs)  & $T_{1}^{\text{s}}[\text{II}]$  (μs) & $T_{2\text{R}}^{\text{s}}[\text{I}]$ (μs) & $T_{2\text{R}}^{\text{s}}[\text{II}]$ (μs) \\
      \hline
      \\[-10pt]
      Device 1 & 5.9 & 0.15 & 0.15 & 2.4 & 4.5 
      & 64 & 73 & 0.47 & 0.21 \\ 
      Device 2 & 7.0 & 0.30 & 0.30 & 3.5 & 7.8 
      & 38 & 44 & 0.25 & 0.18 \\
      Device 3 & 11 & 0.36 & 0.36 & 2.5 & 3.8 
      & 87 & - & 3.90 & - \\
      Device 4 & 8.5 & 0.48 & 0.48 & 2.5 & 5.0 
      & 15 & 8 & 1.40 & 0.35 \\
    \end{tabular}
  \end{ruledtabular}
\end{table*}

\begin{figure}[t]
\includegraphics[width=\columnwidth]{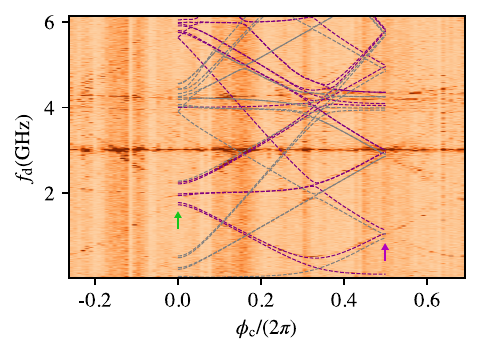}
\protect\caption{\label{fig:figure4}Two-tone spectroscopy of Device 1 as a function of the drive frequency $f_{\text{d}}$ and common-mode flux $\phi_\text{c}$ at $\phi_\text{d} = 0$. Transition frequencies obtained from diagonalizing the protomon Hamiltonian [Eq.~(\ref{eq:fm:ham})] are fit to the spectroscopy data, with circuit parameters $E_\text{J} / h =$ 5.9 GHz, $E_\text{L} / h =$ 0.15 GHz, $E_\text{LS} / h =$ 0.15 GHz, $E_\text{CJ} / h =$ 2.4 GHz, and $E_\text{C} / h = 4.5$ GHz. Frequencies of transitions from the ground state are indicated by dashed purple lines, and those from the first excited state are indicated by dashed gray lines. The flux sweet spots I and II are located at $(\phi_{\text c}/2\pi,\phi_{\text d}/2\pi) = (0, 0)$ (green arrow), and $(\phi_{\text c}/2\pi,\phi_{\text d}/2\pi) = (0.5, 0)$ (purple arrow), respectively. }
\end{figure}

The external flux is controlled by both on-chip flux bias lines (purple regions in Fig.~\ref{fig:figure3}) and a global external magnet. The relative contributions to $\phi_{\text c}$ and $\phi_{\text d}$ are calibrated using cavity transmission data as a function of each source of flux control. Changes in the external flux lead to variations in the interaction between the qubit and the cavity, causing the dispersive shift of the cavity to vary~\cite{Zhu2013}. Identifying the periodic features in the dispersive shifts of the cavity frequency allows for calibrating  $\phi_{\text c}$ and $\phi_{\text d}$ to the applied DC voltage of flux bias line and the global magnet~\cite{Smith2022}. Further details of the flux calibration can be found in Appendix \ref{append:fluxcal}. In order to identify sweet spots of the qubit, we perform two-tone spectroscopy~\cite{Wallraff2007} as a function of $\phi_{\text c}$ (at $\phi_{\text d} = 0$), with the spectrum of Device 1 shown in Fig.~\ref{fig:figure4}. In this measurement, we sweep the frequency of a qubit drive tone ${f_\text{d}}$ applied to the flux bias line while monitoring transmission at the cavity probe frequency. Since the cavity frequency is flux-dependent, we adjust the probe frequency at each flux value to correspond to the minimum in transmission amplitude. We fit the energy spectrum with \texttt{scQubits}~\cite{Groszkowski2021, Chitta2022}, which numerically diagonalizes the full Hamiltonian. The circuit parameters obtained from the fitting each qubit's measured spectrum are summarized in Table~\ref{table:devices_coherences}.

We bias the qubit to sweet spots I and II. Figure \ref{fig:figure4} illustrates sweet spots I (green arrow) and II (purple arrow), showing the two-tone spectroscopy data as a function of $\phi_{\text c}$ (at $\phi_{\text d} = 0$). At the sweet spots, time-domain measurements are conducted to determine the coherence times of the qubits. Additional information about the experimental setup can be found in Appendix \ref{append:experimentalsetup}.

The ability to measure coherence times depends on accurately determining the occupation probabilities of specific quantum states. 
With standard dispersive measurement, we are able to readout the two logical states. Due to the relatively small differences in dispersive shifts, it is challenging to distinguish the logical states from their adjacent levels. 
For example, at sweet spot II, it is particularly difficult to differentiate between the logical state $|0_\text{L}\rangle$ and its adjacent eigenstate that is outside of the logical subspace.
This complication in readout leads to a nonstandard way to characterize the coherence time of the protomon: instead of measuring the probability in logical state $|0_\text{L}\rangle$, we measure the total probability in a subspace $\mathcal{G}$, spanned by $|0_\text{L}\rangle$ and its adjacent levels that cannot be distinguished.

To measure the decay time, we begin by preparing the logical state $|1_\text{L}\rangle$ through selective driving of the logical 0-1 transition. We then monitor the subsequent increase in probability within the subspace $\mathcal{G}$. By fitting this probability change to an exponential decay, we extract the time constant $T_1^\text{s}$, where the superscript ``s'' denotes the subspace. 
To characterize the qubit's dephasing time, we perform the standard Ramsey experiment, tracking the occupation probability within the subspace $\mathcal{G}$ over time. By analyzing the probability decay, we extract the dephasing time constant $T_{2\text{R}}^{\text{s}}$. For a detailed discussion of both $T_1^\text{s}$ and $T_{2\text{R}}^{\text{s}}$, refer to Appendix~\ref{sec: noise}.
Table~\ref{table:devices_coherences} presents the coherence times measured at various sweet spots for protomon devices with different circuit parameters.
Compared to the theoretical predictions, the measured coherence times are relatively short. 
Although the experimental data cannot clearly identify a dominant noise channel, we suspect that \(T_1^\text{s}\) may be limited by capacitive and inductive losses, while \(T_{2\text{R}}^\text{s}\) is likely constrained by flux noise and phase slips.

\section{Conclusion}
\label{sec:conclusion}

In conclusion, we introduced the \textit{protomon}, a multimode qubit based on a fluxonium molecule circuit. We identified three operating points where the qubit subspaces exhibit both disjoint support and first-order insensitivity to flux noise. These theoretical predictions were complemented by the experimental demonstration of the protomon qubit. Four protomon qubits were designed and fabricated, all operating within the desired parameter regime. For each qubit, we precisely calibrated the circuit by tuning both the common and differential flux using the readout resonator's response and two-tone spectroscopy. Time-domain measurements were performed, with measured coherence times shorter than theoretical predictions. The underlying mechanisms for this discrepancy are not fully understood, and we speculate that $T_1$ might be limited by capacitive and inductive losses, while $T_2$ could be constrained by phase slips and flux noise. In future studies, it will be worthwhile to investigate the limiting factors and improve the coherence times of the protomon qubit.

\title{Acknowledgments}
\begin{acknowledgments}
\textit{Acknowledgments---} This research was sponsored by the Army Research Office grants HIPS W911NF1910016 and GASP W911NF2310101. S. Sussman was supported by the Department of Defense (DoD) through the National Defense Science \& Engineering Graduate Fellowship (NDSEG) Program. The devices are fabricated in the PRISM Cleanroom at Princeton University.

Princeton University Professor Andrew A. Houck is also a consultant for Quantum Circuits Incorporated (QCI). Due to his income from QCI, Princeton University has a management plan in place to mitigate a potential conflict of interest that could affect the design, conduct and reporting of this research.
\end{acknowledgments}

\appendix

\section{Quantization of the protomon circuit}
\subsection{Derivation of the full-circuit Lagrangian and Hamiltonian}
\label{app:full_ham}
Here, we derive the Hamiltonian of the full circuit shown in Fig.~\ref{fig:figure1}(a). 
In terms of node variables, the Lagrangian of the circuit is expressed as
\begin{equation*}
\begin{split}
    \mathcal{L} =&\, \dfrac{1}{2}C_\text{J}\phi_0^2 \left[(\dot{\varphi_3}-\dot{\varphi_2})^2 + (\dot{\varphi_4}-\dot{\varphi_1})^2  \right] +  \dfrac{1}{2}C\phi_0^2 (\dot{\varphi_3}-\dot{\varphi_1})^2 \\
    & - \frac{1}{2}E_\text{L}\left[ (\varphi_2 -\varphi_1)^2 + (\varphi_4 -\varphi_3)^2\right] - \frac{1}{2}E_\text{Ls}(\varphi_3-\varphi_1)^2 \\
     & + E_\text{J} \left[ \cos(\varphi_3 - \varphi_2 + \phi_\text{1}) + \cos(\varphi_1 - \varphi_4 + \phi_\text{2}) \right].
     \end{split}
\end{equation*}
Reduced external fluxes are defined as $\phi_\text{1}=\Phi_\text{1}/\phi_0$ and $\phi_\text{2}=\Phi_\text{2}/\phi_0$, with $\phi_0=\hbar/2e$.
To simplify the Lagrangian such that most degrees of freedom are decoupled, we introduce a new set of variables:
\begin{equation*}
    \left(
        \begin{array}{c}
             \phi\\ \theta \\ \zeta \\ \Sigma
        \end{array}
    \right) =\dfrac{1}{2}
    \left(
    \begin{array}{cccc}
         1 & -1 & 1 & -1\\
         -1 & -1 & 1 & 1 \\
         -1 &  0& 1 & 0\\
         1& 1 &1 &1 
    \end{array}
    \right)
    \left(
        \begin{array}{c}
             \varphi_1 \\ \varphi_2 \\ \varphi_3 \\ \varphi_4
        \end{array}
    \right),
\end{equation*}
with the inverse transformation 
\begin{equation*}
    \left(
        \begin{array}{c}
            \varphi_1 \\ \varphi_2 \\ \varphi_3 \\ \varphi_4
        \end{array}
    \right) =\dfrac{1}{2}
    \left(
    \begin{array}{cccc}
         1 & 0 & -1 & 1  \\
         -1 & -2 & 1 & 1 \\
         1 &  0 & 1 &  1  \\
         -1 & 2 & -1 & 1 
    \end{array}
    \right)
    \left(
        \begin{array}{c}
             \phi\\ \theta \\ \zeta \\ \Sigma
        \end{array}
    \right).
\end{equation*}
Applying the Legendre transformation and canonical quantization, one obtains the circuit Hamiltonian in Eq.~\eqref{eq:fm:ham}.

\subsection{Derivation of the reduced Hamiltonian from the Schrieffer-Wolff transformation}\label{app:reduced_ham}
Here, we derived the reduced Hamiltonian in Eq.~\eqref{eq:fm:heff} by employing the Schrieffer--Wolff transformation. 
Following Ref.~\onlinecite{cohen1998atom}, we first rewrite the full circuit Hamiltonian as
\begin{equation}
    \hat{\mathcal{H}} = \hat{\mathcal{H}}_{\phi,\theta} + \hat{\mathcal{H}}_\zeta + \hat{V},
\end{equation}
with the following notations
\begin{equation}
\begin{split}
      \hat{\mathcal{H}}_{\phi,\theta}  = &2 E_\text{CJ} (\hat{n}_\phi^2 + \hat{n}_\theta^2) + E_\text{L} \left(\hat{\phi}^2 + \hat{\theta}^2 \right) \\
     &-2E_\text{J}\cos(\hat{\phi}+\phi_\text{c})\cos(\hat{\theta}+\phi_\text{d}) ,\\
     \hat{\mathcal{H}}_\zeta = & 4 E_\text{C}\hat{n}_\zeta^2 + \left(E_\text{L} 
     + \dfrac{1}{2}E_\text{Ls}\right)\hat{\zeta}^2 , \\
      \hat{V}  = &-2\hat{\theta}\hat{\zeta}.
\end{split}
\end{equation}
The energy spectrum of $\hat{\mathcal{H}}_0$ consists of manifolds $\mathcal{E}_\alpha^0$, $\mathcal{E}_\beta^0$, \dots, with the subscript denotes the number of excitations in $\hat{\mathcal{H}}_\zeta$. Within each manifold, the spectrum is identical to the one of $\hat{\mathcal{H}}_{\phi,\theta}$. The presence of the interaction term $\hat{V}$ perturbs the energy spectrum, and an effective Hamiltonian in the manifold $\alpha$ can be derived from second-order perturbation theory.
To simplify notations, we define the eigenvalues and eigenvectors of the above Hamiltonians following
\begin{equation}
    \hat{\mathcal{H}}_{\phi,\theta} | i\rangle = E_i |i\rangle,  \qquad
    \hat{\mathcal{H}}_\zeta |\alpha\rangle = E_\alpha |\alpha \rangle. 
\end{equation}
In the product basis formed by the eigenstates of $\hat{\mathcal{H}}_{\phi,\theta}$ and $\hat{\mathcal{H}}_\zeta$, the matrix element of the effective Hamiltonian $\hat{\mathcal{H}}_\text{eff}$ can be calculated from
\begin{equation}\label{eq:app:swt}
    \begin{split}
        \langle i | \hat{\mathcal{H}}_\text{eff} - \hat{\mathcal{H}}_0 | j \rangle =&\, \langle i, \alpha | \hat{V} | j,\alpha\rangle \\
        & +\dfrac{1}{2}\sum_k\sum_{\gamma\neq\alpha} \langle i,\alpha| \hat{V} |k,\gamma\rangle \langle k, \gamma | \hat{V} | j,\alpha \rangle \\
        & \times\left(  \dfrac{1}{E_{i\alpha}-E_{k\gamma}} + \dfrac{1}{E_{j\alpha}-E_{k\gamma}} \right), 
    \end{split}
\end{equation}
with the energy $E_{i\alpha} = E_i + E_\alpha$. 
Since $\langle i, \alpha | \hat{V} | j,\alpha\rangle \propto \langle\alpha | \hat{\zeta} | \alpha\rangle = 0$, the first-order perturbation vanishes. 
To evaluate the second-order contribution, we derive the matrix element
\begin{equation}
\begin{split}
    \langle i,\alpha | \hat{V} | k,\gamma\rangle & = -2E_\text{L} \langle i | \hat{\theta} | k\rangle\langle \alpha | \hat{\zeta} | \gamma \rangle \nonumber\\
    & = -2 E_\text{L}\langle i | \hat{\theta} | k\rangle \zeta_\text{zpf}\sqrt{\gamma}\delta_{0,\gamma-1},
\end{split}
\end{equation}
with zero-point fluctuation  $\zeta_\text{zpf}=[4\hbar^2 E_\text{C}/(E_\text{L}+E_\text{Ls}/2)]^{1/4}$. 
The last equality assumes that the $\zeta$ mode is in the ground state at low temperature. 
Together with the approximation $E_{i\alpha},E_{j\alpha} \ll E_{k\gamma}$, Eq.~\eqref{eq:app:swt} reduces to 
\begin{equation*}
    \langle i | \hat{\mathcal{H}}_\text{eff} - \hat{\mathcal{H}}_0 | j \rangle = -\dfrac{2E_\text{L}^2}{2E_\text{L}+E_\text{Ls}}\langle i |\hat{\theta}^2 |j\rangle,
\end{equation*}
from which one obtains 
\begin{equation}
\begin{split}
    \hat{\mathcal{H}}_\text{eff} =&\, \hat{\mathcal{H}}_0 -\dfrac{2E_\text{L}^2}{2E_\text{L}+E_\text{Ls}}\hat{\theta}^2, \nonumber\\
    =&\,  2 E_\text{CJ} \hat{n}_\theta^2  +2 E_\text{CJ} \hat{n}_\phi^2 + \dfrac{E_\text{L}E_\text{Ls}}{2E_\text{L} + E_\text{Ls} } \hat{\theta}^2  + E_\text{L} \hat{\phi} ^2 \\
    &\, - 2 E_\text{J}\cos(\hat{\phi}+\phi_\text{c})\cos(\hat{\theta}+\phi_\text{d}). 
\end{split}
\end{equation}

\subsection{Characterization of disorders in circuit parameters}\label{app:disorder}

The Hamiltonian in Eq.~\eqref{eq:fm:ham} is obtained for a symmetric circuit with identical Josephson junctions and superinductors. In practice, disorder in circuit parameters is inevitable during fabrication. Here, we derive the correction to Eq.~\eqref{eq:fm:ham} in the presence of disorder. 

First, we define the averages and disorder for capacitances, inductances and junction energies as 
\begin{equation*}
    \begin{aligned}
    & C_\text{J} = (C_\text{J1} + C_\text{J2}) / 2, \\
    & L = (L_1 + L_2) / 2, \\
    & E_\text{J} = (E_\text{J1} + E_\text{J2}) / 2,
    \end{aligned}
    \qquad
    \begin{aligned}
    & \mathrm{d} C_\text{J} = (C_\text{J1} - C_\text{J2}) / 2, \\
    & \mathrm{d} L = (L_1 - L_2) / 2, \\
    & \mathrm{d} E_\text{J} = (E_\text{J1} - E_\text{J2}) / 2,
    \end{aligned}
\end{equation*}
with the inverse 
\begin{equation*}
    \begin{aligned}
    & C_\text{J1} = C_\text{J}(1-\mathrm{d} C_\text{J}), \\
    & L_1 = L(1-\mathrm{d} L), \\
    & E_\text{J1} = E_\text{J}(1-\mathrm{d} E_\text{J}),
    \end{aligned}
    \qquad
    \begin{aligned}
    & C_\text{J2} = C_\text{J}(1+\mathrm{d} C_\text{J}), \\
    & L_2 = L(1+\mathrm{d} L), \\
    & E_\text{J2} = E_\text{J}(1+\mathrm{d} E_\text{J}). 
    \end{aligned}
\end{equation*}
If we consider $\mathrm{d} C_\text{J} = \mathrm{d} E_\text{J}= \mathrm{d} A$ as the disorder in junction area, then we have $E_\text{CJ1}E_\text{J1} = E_\text{CJ2}E_\text{J2}$, which reflects that the junction plasma frequency is fixed by oxidation. 

With the above definition, we derive the circuit Hamiltonian in the presence of disorder 
\begin{equation}
    \begin{split}
        \hat{\mathcal{H}} = \,
        & 2 E_\text{CJ}(1-\mathrm{d} C_\text{J}^2)^{-1} (\hat{n}_\phi^2 + \hat{n}_\theta^2) + 4 E_\text{C}\hat{n}_\zeta^2 \\ 
        & + E_\text{L} (1-\mathrm{d} L^2)^{-1} \left[\hat{\phi}^2 + (\hat{\theta}-\hat{\zeta})^2 \right]+ \dfrac{1}{2}E_\text{Ls}\hat{\zeta}^2 \\
        & -2E_\text{J}\cos(\hat{\phi}+\phi_\text{c})\cos(\hat{\theta}+\phi_\text{d}) \\
        & + 4 E_\text{CJ} (1-\mathrm{d} C_\text{J}^2)^{-1}\mathrm{d} C_\text{J}\hat{n}_\theta \hat{n}_\phi\\
        & + 2 E_\text{L}(1-\mathrm{d} L^2)^{-1}  \mathrm{d} L \hat{\phi} (\hat{\theta}-\hat{\zeta})\\
        & -2E_\text{J} \mathrm{d} E_\text{J} \sin(\hat{\phi} + \phi_\text{c})\sin(\hat{\theta} + \phi_\text{d}).
    \end{split}
\end{equation}
Employing a leading-order expansion in the circuit disorder, we cast the Hamiltonian into the form
\begin{equation}
\begin{split}
    \hat{\mathcal{H}} \approx & \hat{\mathcal{H}}_\text{sym} + 4 E_\text{CJ}\mathrm{d} C_\text{J}\hat{n}_\theta \hat{n}_\phi + 2 E_\text{L} \mathrm{d} L \hat{\phi} (\hat{\theta}-\hat{\zeta}) \\
    & -2E_\text{J} \mathrm{d} E_\text{J} \sin(\hat{\phi} + \phi_\text{c})\sin(\hat{\theta} + \phi_\text{d}),
\end{split}
\end{equation}
with the symmetric Hamiltonian $\hat{\mathcal{H}}_\text{sym}$ from Eq.~\eqref{eq:fm:ham}.

\subsection{Analysis of sweet spot II with the hopping model}\label{app:hopping}
\begin{figure}[t]
\includegraphics[width=\columnwidth]{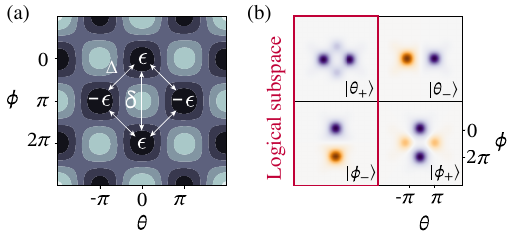}
\protect\caption{\label{fig:hopping} 
(a) Schematics representing  the four-site hopping model at sweet spot II. (b) The wavefunctions of the lowest four eigenstates at sweet spot II in the order: top left, top right, bottom left, and bottom right. 
}
\end{figure}

In order to gain more insight into the structure of the wavefunctions for double sweet spot II, we adopt an effective hopping model. For the inversion-symmetry point II, the schematic of the corresponding four-site hopping model is shown in Fig.~\ref{fig:hopping}. 
Denoting the localized states in the up, down, left and right wells as $|\text{u}\rangle$, $|\text{d}\rangle$, $|\text{l}\rangle$ and $|\text{r}\rangle$, respectively, we obtain the following effective model Hamiltonian
\begin{equation}
\begin{split}
    \hat{H}_\text{hop} = &\, \epsilon \left( |\text{u}\rangle \langle \text{u}| + |\text{d}\rangle \langle \text{d}| \right) -\epsilon \left(|\text{l}\rangle \langle \text{l}| + |\text{r}\rangle \langle \text{r}|\right) \\
     &- \Delta \left( |\text{u}\rangle \langle \text{l}| + |\text{u}\rangle \langle \text{r}| + |\text{d}\rangle \langle \text{l}|+ |\text{d}\rangle \langle \text{r}| +\text{h.c.}\right)
     \\
     & - \delta \left( |\text{u}\rangle \langle \text{d}| + \text{h.c.}  \right).
    \end{split}
\end{equation}
Here, $\pm\epsilon$ denotes the onsite energy associated with the localized states in horizontal and vertical wells, $\Delta$ describes the nearest-neighbor hopping amplitude, and $\delta$ characterizes the next-nearest-neighbor hopping amplitude. The relative magnitudes of these parameters obey $\epsilon \gg \Delta \gg \delta>0$. 
The exact spectrum of the Hamiltonian is calculated and shown in Table~\ref{tab:hop}.

\begin{table}[tb]
    \caption{\label{tab:hop}
    Eigenlevels and wavefunctions obtained from the four-site hopping model.}
    \begin{ruledtabular}
        \begin{tabular}{ccc}
            Eigenstate & Energy & Wavefunction \\
            \hline \\[-10pt]
            $|\theta_+\rangle$ & $-\epsilon-2\Delta^2/\epsilon$ & $( \Delta/\epsilon, \Delta/\epsilon ,1, 1 )$ \\
            $|\theta_-\rangle$ & $-\epsilon$ & $( 0,0 ,-1, 1 )$ \\
            $|\phi_-\rangle$ & $\epsilon+\delta$ & $(1,-1,0,0)$ \\
            $|\phi_+\rangle$ & $\epsilon-\delta+2\Delta^2/\epsilon$ & $(1,1 -\Delta/\epsilon, -\Delta/\epsilon)$ \\
        \end{tabular}
    \end{ruledtabular}
\end{table}

The calculated eigenvectors fully capture the parities of the wavefunctions and the weight in each potential well. 
The symmetric state, e.g., $|\theta_+\rangle$, has dominant weight in wells along the $\theta$ direction, while minor in wells along the $\phi$ direction. The relative weight is governed by the ratio of the onsite energy and the nearest-neighbor tunneling amplitude $\epsilon/\Delta$. 
In contrast, the anti-symmetric state, e.g., $|\phi_-\rangle$, only occupies wells along the $\phi$ direction. 
The odd parity of the wavefunction leads to destructive interference between the two paths starting from the upper and lower sites to the horizontal site along the $\theta$ direction, and thus the absence of population there. 
Which one of the two states $|\phi_-\rangle$ and $|\phi_+\rangle$ has lower energy is parameter dependent. When $\delta > \Delta^2/\epsilon$, the symmetric state has a lower energy compared with the anti-symmetric one. This regime may be achieved, for example, by increasing the inductive energies of the circuit, which effectively enhances the hopping amplitude $\delta$ between next nearest neighbors. 
Specifically, for devices 1, 2, and 3, we have \( |0\rangle \equiv |\theta_+\rangle \), \( |1\rangle \equiv |\theta_-\rangle \), \( |2\rangle \equiv |\phi_-\rangle \), \( |3\rangle \equiv |\phi_+\rangle \); and for Device 4, we have \( |0\rangle \equiv |\theta_+\rangle \), \( |1\rangle \equiv |\theta_-\rangle \), \( |2\rangle \equiv |\phi_+\rangle \), \( |3\rangle \equiv |\phi_-\rangle \).

\section{\label{sec: noise}Decoherence channels and coherence times}

In this section, we detail the potential decoherence channels for both depolarization and pure dephasing. Subsequently, we explore the appropriate metrics for coherence times in the presence of unresolved nearby states. 

\subsection{Depolarization loss mechanisms}
\label{append:noiseT1}
The depolarization rate for spontaneous transitions from $\ket{i}$ to $\ket{j}$ due to noise source $\lambda$ can be estimated using Fermi's golden rule~\cite{Clerk2010},
\begin{equation}
\label{equation:Fermi}
\Gamma_{1,ij}^\lambda = \frac{1}{\hbar^2}|\bra{i}\hat{G}_\lambda\ket{j}|^2 S_\lambda(\omega_{ij}).
\end{equation}
Here, $\hat{G}_\lambda$ is the qubit operator which couples to the noise, and $S_\lambda(\omega_{ij})$ is the noise spectral density evaluated at the transition frequency $\omega_{ij}=\omega_i - \omega_j$. We next detail the dominant sources of depolarization in the protomon.

\paragraph{Capacitive loss}
Capacitive loss is caused by dissipation of electromagnetic energy in dielectrics, materials interfaces, junction tunnel barriers and the substrate~\cite{Martinis2017, Oliver2019, Gambetta2016}. Both small Josephson junctions in the protomon have a small capacitance across them, and thus the noise operator $\hat{G}_\text{cap}$ takes the form $\hat{G}_\text{cap}$ = $e$($\hat{n}_\phi\pm\hat{n}_\theta$). We assume, for simplicity, that capacitive noise sources coupled to each mode are independent. The noise spectral density of capacitive loss is 
\begin{equation}
\label{equation:capspectraldensity}
S_\text{cap}(\omega) = \frac{\hbar}{C_\text{J} Q_\text{cap}(\omega)}\left(\coth{\frac{\hbar\omega}{2k_\text{B}T}}+1\right),
\end{equation}
where $C_\text{J}$ is the junction capacitance, $Q_\text{cap}$ is the dielectric quality factor, $T$ is the temperature and $k_\text{B}$ is Boltzmann's constant. We assume the dielectric quality factor follows \cite{nguyenHighCoherenceFluxoniumQubit2019}
\begin{equation}
    Q_\text{cap} (\omega) = 
    Q_\text{cap}(2\pi \times 6\text{ GHz}) 
    \left (
    \frac{2 \pi \times 6\text{ GHz}}{|\omega|}
    \right )^{0.7},
\end{equation}
where $Q_\text{cap}(2\pi \times 6\text{ GHz})$ is the reference inductive quality factor at frequency $\omega = 2\pi \times 6\text{ GHz}$.

\paragraph{Inductive loss}
The noise operator for inductive loss is the phase across each inductor: the protomon is a multi-mode circuit with three inductors, and therefore has three noise operators to characterize inductive loss, given by $\hat{G}_\text{ind}$ = $\phi_0$($\pm\hat{\phi}+\hat{\theta}-\hat{\zeta}$) and $\phi_0\hat{\zeta}$. Since the protomon's inductors comprise arrays of Josephson junctions, inductive loss is thought to originate in quasiparticle tunneling across array junctions~\cite{Vool2014}. Assuming that the frequency dependence of the real part of the admittance for inductive loss mimics quasiparticle losses~\cite{Pop2014, Catelani2011}, the noise spectral density for inductive loss is
\begin{equation}
\label{equation:indspectraldensity}
S_\text{ind}(\omega) = \frac{\hbar}{L Q_\text{ind}(\omega)}\left(\coth{\frac{\hbar\omega}{2k_\text{B}T}}+1\right), 
\end{equation}
where we assume identical inductors, $E_{\text L}$ = $E_{\text{Ls}}$, and $L$ is the inductance of each array. $Q_\text{ind}(\omega)$ is the frequency dependent inductive quality factor. We assume the dielectric quality factor follows \cite{nguyenHighCoherenceFluxoniumQubit2019}
\begin{equation}
\begin{aligned}
    Q_\text{ind} (\omega) &= 
    Q_\text{ind}(2\pi \times 0.5\text{ GHz}) \\
    &\times\frac{K_0 \left( \frac{h \times 0.5 \text{ GHz}}{2 k_\text{B} T} \right) \sinh \left( \frac{h \times 0.5 \text{ GHz}}{2 k_\text{B} T} \right)  }
    {K_0 \left( \frac{\hbar |\omega|}{2 k_\text{B} T} \right) 
    \sinh \left( \frac{\hbar |\omega| }{2 k_\text{B} T} \right) }
\end{aligned}
\end{equation}
where $Q_\text{ind}(2\pi \times 0.5\text{ GHz})$ is the reference inductive quality factor at frequency $\omega = 2\pi \times 0.5\text{ GHz}$.

\paragraph{Quasiparticle loss}
Non-equilibirum quasiparticles with energies above the superconducting gap contribute to qubit relaxation, where single electrons tunneling across junctions can change the charge parity of the qubit, as well as create parallel dissipative conducting channels \cite{Serniak2018, Pop2014}. The quasiparticle noise operator couples to the phase across single Josephson junctions via $\hat{G}_\text{qp}$ = 2$\phi_0\sin({\hat{\phi}\pm\hat{\theta}}/2)$. The quasiparticle noise spectral density is given by
\begin{equation}
\label{equation:qpspectraldensity}
S_\text{qp}(\omega) = \hbar\omega \mathrm{Re}(Y_\text{qp})(\omega)\left(\coth\frac{\hbar\omega}{2k_\text{B}T}+1\right)
\end{equation}
where $\mathrm{Re}(Y_\text{qp})$ is the real part of the complex admittance, and is given by
\begin{equation}
\label{equation:quasiadmittance}
\begin{aligned}
\mathrm{Re}(Y_\text{qp}) &= \sqrt{\frac{2}{\pi}}\frac{8E_\text{J}}{R_\text{K}\Delta}\left(\frac{2\Delta}{\hbar\omega}\right)^{3/2}x_\text{qp}\sqrt{\frac{\hbar\omega}{2k_\text{B}T}}\\
&\times K_0\left(\frac{\hbar|\omega|}{2k_\text{B}T}\right)\sinh\left(\frac{\hbar\omega}{2k_\text{B}T}\right)
\end{aligned}
\end{equation}
where $\Delta$ is the superconducting gap, $R_\text{K} = h/e^2$ is the superconducting resistance quantum and $x_\text{qp}$ is the quasiparticle density normalized by the Cooper pair density.

\paragraph{Other possible noise channels}
Besides capacitive loss and inductive loss, other noise sources might contribute to depolarization time. 
For example, the qubit coupling to the readout resonator could lead to spontaneous emission of photons, known as the Purcell effect~\cite{Houck2008}. 
Our numerical evidences (not shown here) suggest that these channels are subdominant for our theoretically proposed parameters.

\begin{table}[tb]
    \caption{\label{tab:noise_param}
    Noise parameters used for the theoretical estimation of coherence times.}
    \begin{ruledtabular}
        \begin{tabular}{ccc}
            Parameter & Value & Reference \\
            \hline \\[-10pt]
            $T$ & $50$ mK & - \\
            $Q_\text{cap}(2\pi \times 6\text{ GHz})$ & $10^6$ & \cite{Wang2015} \\
            $Q_\text{ind}(2\pi \times 0.5\text{ GHz})$ & $5 \times 10^8$ & \cite{Pop2014} \\
            $\Delta$ & $3.4 \times 10^{-4}$ eV& \cite{kittel} \\
            $x_\text{qp}$ & $10^{-8}$ & \cite{nguyenHighCoherenceFluxoniumQubit2019} \\
        \end{tabular}
    \end{ruledtabular}
\end{table}

\paragraph{Noise parameters}
The theoretically estimated coherence times as shown in Fig.~\ref{fig:coherence} are based on assuming parameters in Table \ref{tab:noise_param}, where the quoted numbers are found typical in previous literature.

\subsection{Pure dephasing mechanisms}
\label{append:noiseTphi}
Here we detail the noise channels that may contribute to protomon's pure dephasing time. 

\paragraph{$1/f$ flux noise}
Qubits with superconducting loops like protomon are subjective to low-frequency flux noise~\cite{KochR2007,Wang2015}, which is attributed to surface defects such as adsorbed molecular O${}_2$~\cite{Bialczak2007,Kumar2016}. 
The first-order contribution of dephasing rate between states $\ket{i}$ and $\ket{j}$ due to the $1/f$ flux noise is
\begin{equation}
\label{equation:dephasingfirstorder}
\Gamma_{\phi,ij}^\text{flux} = \sum_{\lambda\in \{\phi_{\text c}, \phi_{\text d}\}} A_{\lambda} \sqrt{|\ln 2\omega_{\text{ir}}t|}\left|\pdv{E_{ij}}{\lambda}\right|
\end{equation}
where $A_\lambda$ is the amplitude of the noise channel, $\omega_{\text{ir}}$ is a low-frequency cutoff, $t$ is the Ramsey experiment time, and $\Delta E$ is the transition energy between the two logical states. At sweet spots, the first-order contribution vanishes, and the pure dephasing rate is dominated by second-order contribution.

\paragraph{Other possible noise channels}
We expect photon shot noise \cite{Groszkowski2018} to have a negligible impact on coherence times since the $\zeta$ mode energy is substantially higher than the thermal energy in the targeted parameter regime. In addition, we estimate the dephasing due to critical-current noise~\cite{vanharlingen2004} (not shown here) and conclude that it is unlikely the dominant contribution to the total dephasing rate.

\subsection{\label{sec: noise-subspace}Decoherence dynamics of multi-level systems}

As discussed in Sec.~\ref{section:theory}, the logical subspace of the protomon is not necessarily the lowest two eigenstates. While coherent transitions to non-logical states can be avoided with selective control pulses, their contribution to decoherence dynamics (leakage) is non-negligible. In particular, at the operation points, these non-logical states become near-degenerate with the logical ones. For example, at sweet spot I, the eigenstate $\ket{2}$ is near-degenerate with $\ket{1_L} \equiv \ket{1}$. At sweet spot II, the spectrum features two doublets: $\{\ket{0_L} \equiv \ket{0}, \ket{1}\}$ and $\{\ket{1_L} \equiv \ket{2}, \ket{3}\}$. With our circuit parameters, those transition energies are smaller compared to $k_\text{B} T$, making leakage a possible source of both depolarization and dephasing.

To account for leakage, we model the decoherence dynamics with the following master equation
\begin{equation}
    \begin{aligned}
        \frac{\mathrm{d}\hat{\rho}}{\mathrm{d}t} = -i [\hat{\mathcal{H}}, \hat{\rho}] 
        +  \sum_{i\neq j} \Gamma_{1,ij} \mathcal{D}[\ket{i}\bra{j}] \hat{\rho},
    \end{aligned}
\end{equation}
with $\Gamma_{1,ij} \equiv \sum_\lambda \Gamma_{1,ij}^\lambda$.
Here, we neglect pure dephasing and focus on the contribution from leakage transitions for simplicity.
The decoherence dynamics determined by the above equation is generally not a single-exponential function of time. Specifically, it exhibits a multi-exponential decay with different rates and asymptotes in the long-time limit. Therefore, it is insufficient to characterize the full decoherence dynamics with just one time constant. However, in the short-time limit, it is possible to derive the initial incoherent transition rate of the occupation probability for a certain state, which can serve as a figure-of-merit for the decoherence properties of the qubit.

To characterize the incoherent transition from $\ket{n}$, the initial decay rate of $\bra{n}\hat{\rho}(t)\ket{n}$ is 
\begin{equation}
    \begin{aligned}
        \Gamma_{1,n} = \sum_{j \neq n} \Gamma_{1,nj}.
    \end{aligned}
    \label{eq: actual gamma 1}
\end{equation}
Specifically, $\Gamma_{1,0_\text{L}}$ and $\Gamma_{1,1_\text{L}}$ together determine the  depolarization of the logical qubit. 
Without leakage, $\Gamma_{1, 0_\text{L}} + \Gamma_{1,1_\text{L}}$ reduces to the usual definition of the depolarization rate of a two-level system $\Gamma_{1, 0_\text{L}1_\text{L}} + \Gamma_{1,1_\text{L}0_\text{L}}$.
To describe the depolarization-induced dephasing process between $\ket{0_\text{L}}$ and $\ket{1_\text{L}}$, we  calculate the dynamics of $|\bra{0_\text{L}}\hat{\rho}(t)\ket{1_\text{L}}|$ as 
\begin{equation}
    \Gamma_2 = 
    \sum_{j \neq 0_\text{L}} \frac{1}{2} \Gamma_{1, 0_\text{L}j} 
    + \sum_{j \neq 1_\text{L}} \frac{1}{2} \Gamma_{1, 1_\text{L}j}. 
\end{equation}
Without leakage, the above expression reduces to  $(\Gamma_{1, 0_\text{L}1_\text{L}} + \Gamma_{1,1_\text{L}0_\text{L}})/2$.

\subsection{Measurement of inter-subspace coherence times}
Typically, the coherence times are measured by probing the time-dependent occupation probability of a logical state through standard dispersive readout~\cite{gambettaProtocolsOptimalReadout2007, blaisCircuitQuantumElectrodynamics2021}.
However, in our setup, the near-degenerate states introduced in Appendix~\ref{sec: noise-subspace} cannot be distinguished because of their similar dispersive shifts.
For example, we can only measure the occupation probability of the subspace \(\mathcal{G}\), which is spanned by \(|0\rangle\) and its near-degenerate levels.
Specifically, the occupation probability within a subspace \(\mathcal{G}\) is defined as follows:
\begin{equation}
    \begin{aligned}
        P_\mathcal{G}(\rho) = \sum_{ \{ \ket{i_\mathcal{G}} \} } \bra{i_\mathcal{G}} \rho \ket{i_\mathcal{G}},
    \end{aligned}
    \label{eq: PG def}
\end{equation}
where \(\{ \ket{i_\mathcal{G}} \}\) represents the states spanning \(\mathcal{G}\).

\paragraph{Inter-subspace relaxation time $T_1^{\text{s}}$}
In our experiments, we investigate the relaxation dynamics by measuring \(P_\mathcal{G}(\rho)\).
First, we initialize \(\rho(t=0) = \ket{1_\text{L}}\bra{1_\text{L}}\) by applying a \(\pi\) pulse that selectively drives the \( \ket{0_\text{L}} - \ket{1_\text{L}} \) transition while avoiding spurious ones.
This selective excitation is achieved using a weak flux drive, where the amplitude is much smaller than the detuning from other unwanted transitions.
Next, we measure \(P_\mathcal{G}(\rho(t))\) as a function of time to track the relaxation dynamics.
We then fit the measured \(P_\mathcal{G}(\rho(t))\) using \(A + B\exp(- t / T_{1}^{\text{s}} )\), where \(T_{1}^{\text{s}}\) represents the extracted decay time scale.
The fitted values for various qubits and sweet spots are presented in Table~\ref{table:devices_coherences}.

\paragraph{Inter-subspace dephasing time $T_{2R}^\text{s}$}
We perform a standard Ramsey experiment to characterize the dephasing time between \(\ket{0_\text{L}}\) and \(\ket{1_\text{L}}\).
First, we prepare the superposition state \(\ket{0_\text{L}} + \ket{1_\text{L}}\), let it idle for a time \(t\), and then map the state back to \(\ket{0_\text{L}}\).
We then measure \(P_\mathcal{G}(\rho(t))\) as a function of the idling time \(t\) for the final state \(\rho(t)\).
Similar to the relaxation time extraction, we fit the envelope of \(P_\mathcal{G}(\rho(t))\) with an exponential function, yielding the overall timescale \(T_\text{2R}^\text{s}\).
The fitted values for various qubits and sweet spots are presented in Table~\ref{table:devices_coherences}.

\section{Device fabrication}

Devices are fabricated on sapphire substrates coated with 200 nm of sputtered tantalum, chosen for its superior performance in transmon qubits~\cite{Place2021}. Resonators, coplanar waveguides, and ground planes are defined through optical lithography and dry etching. Josephson junctions and inductor arrays are patterned using electron beam lithography. Devices are developed in a MIBK:IPA solution, with an argon ion beam etch performed before deposition. For Dolan junctions, aluminum layers are evaporated at specific angles, and the first layer is oxidized to form tunnel barriers between superconducting islands.

\section{Experimental setup}
\label{append:experimentalsetup}

The devices are wire-bonded to a copper PCB and mounted on a dilution fridge base plate at 17 mK. They are shielded with copper, Eccosorb-coated aluminum, and mu-metal. We update our experimental configuration in various iterations of the devices, transitioning from using a set of AWGs and digitizers to using the Quantum Instrumentation Control Kit (QICK)~\cite{Ding2024,stefanazzi2022qick}. Time domain measurements are performed with the QICK control system for devices 1, 2, and 4. Device 3 is measured using a homodyne detection method.

\section{Flux calibration and spectrum fitting}
\label{append:fluxcal}

Flux control in all devices is achieved via an on-chip flux bias line and an external magnet. The contributions to $\phi_{\text c}$ and $\phi_{\text d}$ are calibrated by first plotting transmission as a function of probe frequency $f$ and magnet bias current, centered around the cavity frequency $f_c$. This data helps select a probe frequency that captures both resonant transmission and vacuum Rabi splitting events. Transmission is mapped against external and on-chip flux bias at this frequency, generating a dataset with periodic features corresponding to single flux quanta along $\phi_{\text c}$ and $\phi_{\text d}$~\cite{Smith2022}.

The external solenoid magnet and on-chip flux bias line control the common ($\phi_{\text c}$) and differential ($\phi_{\text d}$) mode flux. The external magnet, concentric with the qubit chip, mainly affects $\phi_{\text c}$, while the on-chip bias line couples to both $\phi_{\text c}$ and $\phi_{\text d}$. We assume flux is linear with current and express total $\phi_{\text c}$ and $\phi_{\text d}$ as:

\[
\begin{bmatrix}
\phi_{\text c}  \\
\phi_{\text d}
\end{bmatrix} =
\begin{bmatrix}
\alpha_m & \alpha_{l} \\
\beta_m & \beta_{l} 
\end{bmatrix}
\begin{bmatrix}
v_{m}  \\
v_{l}
\end{bmatrix} +
\begin{bmatrix}
\phi_{\delta \text{c}}  \\
\phi_{\delta \text{d}}
\end{bmatrix}.
\]

Here, $v_{m}$ and $v_{l}$ denote the voltages applied to the external magnet and the on-chip flux bias line, respectively, while $\phi_{\delta \text{c}}$ and $\phi_{\delta \text{d}}$ represent the constant flux offsets. Typically, the external magnet's dynamic range is larger, coupling strongly to $\phi_{\text c}$, while the on-chip bias line does not reach a full flux quantum in $\phi_{\text d}$. To parametrize $\phi_{\text d}$, spectra are measured at various flux values along an axis parallel to $\phi_{\text c}$. Fitting a model implemented in \texttt{scQubits} to these spectra allows us to extract the parameters. The fitted two-tone spectroscopy data for Device 1 are shown against the calibrated $\phi_{\text c}$ at $\phi_{\text d}$ = 0 (see Fig.~\ref{fig:figure4}).

\bibliography{main}

\begin{thebibliography}{55}%
\makeatletter
\providecommand \@ifxundefined [1]{%
 \@ifx{#1\undefined}
}%
\providecommand \@ifnum [1]{%
 \ifnum #1\expandafter \@firstoftwo
 \else \expandafter \@secondoftwo
 \fi
}%
\providecommand \@ifx [1]{%
 \ifx #1\expandafter \@firstoftwo
 \else \expandafter \@secondoftwo
 \fi
}%
\providecommand \natexlab [1]{#1}%
\providecommand \enquote  [1]{``#1''}%
\providecommand \bibnamefont  [1]{#1}%
\providecommand \bibfnamefont [1]{#1}%
\providecommand \citenamefont [1]{#1}%
\providecommand \href@noop [0]{\@secondoftwo}%
\providecommand \href [0]{\begingroup \@sanitize@url \@href}%
\providecommand \@href[1]{\@@startlink{#1}\@@href}%
\providecommand \@@href[1]{\endgroup#1\@@endlink}%
\providecommand \@sanitize@url [0]{\catcode `\\12\catcode `\$12\catcode `\&12\catcode `\#12\catcode `\^12\catcode `\_12\catcode `\%12\relax}%
\providecommand \@@startlink[1]{}%
\providecommand \@@endlink[0]{}%
\providecommand \url  [0]{\begingroup\@sanitize@url \@url }%
\providecommand \@url [1]{\endgroup\@href {#1}{\urlprefix }}%
\providecommand \urlprefix  [0]{URL }%
\providecommand \Eprint [0]{\href }%
\providecommand \doibase [0]{https://doi.org/}%
\providecommand \selectlanguage [0]{\@gobble}%
\providecommand \bibinfo  [0]{\@secondoftwo}%
\providecommand \bibfield  [0]{\@secondoftwo}%
\providecommand \translation [1]{[#1]}%
\providecommand \BibitemOpen [0]{}%
\providecommand \bibitemStop [0]{}%
\providecommand \bibitemNoStop [0]{.\EOS\space}%
\providecommand \EOS [0]{\spacefactor3000\relax}%
\providecommand \BibitemShut  [1]{\csname bibitem#1\endcsname}%
\let\auto@bib@innerbib\@empty
\bibitem [{\citenamefont {\relax{Google Quantum AI}}(2023)}]{google2023suppressing}%
  \BibitemOpen
  \bibfield  {author} {\bibinfo {author} {\bibnamefont {\relax{Google Quantum AI}}},\ }\bibfield  {title} {\bibinfo {title} {Suppressing quantum errors by scaling a surface code logical qubit},\ }\href {https://doi.org/10.1038/s41586-022-05434-1} {\bibfield  {journal} {\bibinfo  {journal} {Nature}\ }\textbf {\bibinfo {volume} {614}},\ \bibinfo {pages} {676} (\bibinfo {year} {2023})}\BibitemShut {NoStop}%
\bibitem [{\citenamefont {Clark}\ \emph {et~al.}(2021)\citenamefont {Clark}, \citenamefont {Tinkey}, \citenamefont {Sawyer}, \citenamefont {Meier}, \citenamefont {Burkhardt}, \citenamefont {Seck}, \citenamefont {Shappert}, \citenamefont {Guise}, \citenamefont {Volin}, \citenamefont {Fallek}, \citenamefont {Hayden}, \citenamefont {Rellergert},\ and\ \citenamefont {Brown}}]{PhysRevLett.127.130505}%
  \BibitemOpen
  \bibfield  {author} {\bibinfo {author} {\bibfnamefont {C.~R.}\ \bibnamefont {Clark}}, \bibinfo {author} {\bibfnamefont {H.~N.}\ \bibnamefont {Tinkey}}, \bibinfo {author} {\bibfnamefont {B.~C.}\ \bibnamefont {Sawyer}}, \bibinfo {author} {\bibfnamefont {A.~M.}\ \bibnamefont {Meier}}, \bibinfo {author} {\bibfnamefont {K.~A.}\ \bibnamefont {Burkhardt}}, \bibinfo {author} {\bibfnamefont {C.~M.}\ \bibnamefont {Seck}}, \bibinfo {author} {\bibfnamefont {C.~M.}\ \bibnamefont {Shappert}}, \bibinfo {author} {\bibfnamefont {N.~D.}\ \bibnamefont {Guise}}, \bibinfo {author} {\bibfnamefont {C.~E.}\ \bibnamefont {Volin}}, \bibinfo {author} {\bibfnamefont {S.~D.}\ \bibnamefont {Fallek}}, \bibinfo {author} {\bibfnamefont {H.~T.}\ \bibnamefont {Hayden}}, \bibinfo {author} {\bibfnamefont {W.~G.}\ \bibnamefont {Rellergert}},\ and\ \bibinfo {author} {\bibfnamefont {K.~R.}\ \bibnamefont {Brown}},\ }\bibfield  {title} {\bibinfo {title} {High-fidelity bell-state preparation with $^{40}{\mathrm{ca}}^{+}$ optical qubits},\ }\href
  {https://doi.org/10.1103/PhysRevLett.127.130505} {\bibfield  {journal} {\bibinfo  {journal} {Phys. Rev. Lett.}\ }\textbf {\bibinfo {volume} {127}},\ \bibinfo {pages} {130505} (\bibinfo {year} {2021})}\BibitemShut {NoStop}%
\bibitem [{\citenamefont {Ryan-Anderson}\ \emph {et~al.}(2021)\citenamefont {Ryan-Anderson}, \citenamefont {Bohnet}, \citenamefont {Lee}, \citenamefont {Gresh}, \citenamefont {Hankin}, \citenamefont {Gaebler}, \citenamefont {Francois}, \citenamefont {Chernoguzov}, \citenamefont {Lucchetti}, \citenamefont {Brown}, \citenamefont {Gatterman}, \citenamefont {Halit}, \citenamefont {Gilmore}, \citenamefont {Gerber}, \citenamefont {Neyenhuis}, \citenamefont {Hayes},\ and\ \citenamefont {Stutz}}]{PhysRevX.11.041058}%
  \BibitemOpen
  \bibfield  {author} {\bibinfo {author} {\bibfnamefont {C.}~\bibnamefont {Ryan-Anderson}}, \bibinfo {author} {\bibfnamefont {J.~G.}\ \bibnamefont {Bohnet}}, \bibinfo {author} {\bibfnamefont {K.}~\bibnamefont {Lee}}, \bibinfo {author} {\bibfnamefont {D.}~\bibnamefont {Gresh}}, \bibinfo {author} {\bibfnamefont {A.}~\bibnamefont {Hankin}}, \bibinfo {author} {\bibfnamefont {J.~P.}\ \bibnamefont {Gaebler}}, \bibinfo {author} {\bibfnamefont {D.}~\bibnamefont {Francois}}, \bibinfo {author} {\bibfnamefont {A.}~\bibnamefont {Chernoguzov}}, \bibinfo {author} {\bibfnamefont {D.}~\bibnamefont {Lucchetti}}, \bibinfo {author} {\bibfnamefont {N.~C.}\ \bibnamefont {Brown}}, \bibinfo {author} {\bibfnamefont {T.~M.}\ \bibnamefont {Gatterman}}, \bibinfo {author} {\bibfnamefont {S.~K.}\ \bibnamefont {Halit}}, \bibinfo {author} {\bibfnamefont {K.}~\bibnamefont {Gilmore}}, \bibinfo {author} {\bibfnamefont {J.~A.}\ \bibnamefont {Gerber}}, \bibinfo {author} {\bibfnamefont {B.}~\bibnamefont {Neyenhuis}}, \bibinfo {author}
  {\bibfnamefont {D.}~\bibnamefont {Hayes}},\ and\ \bibinfo {author} {\bibfnamefont {R.~P.}\ \bibnamefont {Stutz}},\ }\bibfield  {title} {\bibinfo {title} {Realization of real-time fault-tolerant quantum error correction},\ }\href {https://doi.org/10.1103/PhysRevX.11.041058} {\bibfield  {journal} {\bibinfo  {journal} {Phys. Rev. X}\ }\textbf {\bibinfo {volume} {11}},\ \bibinfo {pages} {041058} (\bibinfo {year} {2021})}\BibitemShut {NoStop}%
\bibitem [{\citenamefont {Zajac}\ \emph {et~al.}(2021)\citenamefont {Zajac}, \citenamefont {Stehlik}, \citenamefont {Underwood}, \citenamefont {Phung}, \citenamefont {Blair}, \citenamefont {Carnevale}, \citenamefont {Klaus}, \citenamefont {Keefe}, \citenamefont {Carniol}, \citenamefont {Kumph}, \citenamefont {Steffen},\ and\ \citenamefont {Dial}}]{zajac2021spectator}%
  \BibitemOpen
  \bibfield  {author} {\bibinfo {author} {\bibfnamefont {D.~M.}\ \bibnamefont {Zajac}}, \bibinfo {author} {\bibfnamefont {J.}~\bibnamefont {Stehlik}}, \bibinfo {author} {\bibfnamefont {D.~L.}\ \bibnamefont {Underwood}}, \bibinfo {author} {\bibfnamefont {T.}~\bibnamefont {Phung}}, \bibinfo {author} {\bibfnamefont {J.}~\bibnamefont {Blair}}, \bibinfo {author} {\bibfnamefont {S.}~\bibnamefont {Carnevale}}, \bibinfo {author} {\bibfnamefont {D.}~\bibnamefont {Klaus}}, \bibinfo {author} {\bibfnamefont {G.~A.}\ \bibnamefont {Keefe}}, \bibinfo {author} {\bibfnamefont {A.}~\bibnamefont {Carniol}}, \bibinfo {author} {\bibfnamefont {M.}~\bibnamefont {Kumph}}, \bibinfo {author} {\bibfnamefont {M.}~\bibnamefont {Steffen}},\ and\ \bibinfo {author} {\bibfnamefont {O.~E.}\ \bibnamefont {Dial}},\ }\bibfield  {title} {\bibinfo {title} {Spectator errors in tunable coupling architectures},\ }\href {https://arxiv.org/abs/2108.11221} {\bibfield  {journal} {\bibinfo  {journal} {arXiv:2108.11221}\ } (\bibinfo {year}
  {2021})}\BibitemShut {NoStop}%
\bibitem [{\citenamefont {Evered}\ \emph {et~al.}(2023)\citenamefont {Evered}, \citenamefont {Bluvstein}, \citenamefont {Kalinowski}, \citenamefont {Ebadi}, \citenamefont {Manovitz}, \citenamefont {Zhou}, \citenamefont {Li}, \citenamefont {Geim}, \citenamefont {Wang}, \citenamefont {Maskara} \emph {et~al.}}]{evered2023high}%
  \BibitemOpen
  \bibfield  {author} {\bibinfo {author} {\bibfnamefont {S.~J.}\ \bibnamefont {Evered}}, \bibinfo {author} {\bibfnamefont {D.}~\bibnamefont {Bluvstein}}, \bibinfo {author} {\bibfnamefont {M.}~\bibnamefont {Kalinowski}}, \bibinfo {author} {\bibfnamefont {S.}~\bibnamefont {Ebadi}}, \bibinfo {author} {\bibfnamefont {T.}~\bibnamefont {Manovitz}}, \bibinfo {author} {\bibfnamefont {H.}~\bibnamefont {Zhou}}, \bibinfo {author} {\bibfnamefont {S.~H.}\ \bibnamefont {Li}}, \bibinfo {author} {\bibfnamefont {A.~A.}\ \bibnamefont {Geim}}, \bibinfo {author} {\bibfnamefont {T.~T.}\ \bibnamefont {Wang}}, \bibinfo {author} {\bibfnamefont {N.}~\bibnamefont {Maskara}}, \emph {et~al.},\ }\bibfield  {title} {\bibinfo {title} {High-fidelity parallel entangling gates on a neutral-atom quantum computer},\ }\href@noop {} {\bibfield  {journal} {\bibinfo  {journal} {Nature}\ }\textbf {\bibinfo {volume} {622}},\ \bibinfo {pages} {268} (\bibinfo {year} {2023})}\BibitemShut {NoStop}%
\bibitem [{\citenamefont {Knill}(2005)}]{Knill2005}%
  \BibitemOpen
  \bibfield  {author} {\bibinfo {author} {\bibfnamefont {E.}~\bibnamefont {Knill}},\ }\bibfield  {title} {\bibinfo {title} {Quantum computing with realistically noisy devices},\ }\href {https://doi.org/10.1038/nature03350} {\bibfield  {journal} {\bibinfo  {journal} {Nature}\ }\textbf {\bibinfo {volume} {434}},\ \bibinfo {pages} {39} (\bibinfo {year} {2005})}\BibitemShut {NoStop}%
\bibitem [{\citenamefont {Wang}\ \emph {et~al.}(2015)\citenamefont {Wang}, \citenamefont {Shi}, \citenamefont {Hu}, \citenamefont {Han}, \citenamefont {Yu},\ and\ \citenamefont {Wu}}]{Wang2015}%
  \BibitemOpen
  \bibfield  {author} {\bibinfo {author} {\bibfnamefont {H.}~\bibnamefont {Wang}}, \bibinfo {author} {\bibfnamefont {C.}~\bibnamefont {Shi}}, \bibinfo {author} {\bibfnamefont {J.}~\bibnamefont {Hu}}, \bibinfo {author} {\bibfnamefont {S.}~\bibnamefont {Han}}, \bibinfo {author} {\bibfnamefont {C.~C.}\ \bibnamefont {Yu}},\ and\ \bibinfo {author} {\bibfnamefont {R.~Q.}\ \bibnamefont {Wu}},\ }\bibfield  {title} {\bibinfo {title} {Candidate source of flux noise in squids: Adsorbed oxygen molecules},\ }\href {https://doi.org/10.1103/PhysRevLett.115.077002} {\bibfield  {journal} {\bibinfo  {journal} {Phys. Rev. Lett.}\ }\textbf {\bibinfo {volume} {115}},\ \bibinfo {pages} {077002} (\bibinfo {year} {2015})}\BibitemShut {NoStop}%
\bibitem [{\citenamefont {Place}\ \emph {et~al.}(2021)\citenamefont {Place}, \citenamefont {Rodgers}, \citenamefont {Mundada}, \citenamefont {Smitham}, \citenamefont {Fitzpatrick}, \citenamefont {Leng}, \citenamefont {Premkumar}, \citenamefont {Bryon}, \citenamefont {Vrajitoarea}, \citenamefont {Sussman}, \citenamefont {Cheng}, \citenamefont {Madhavan}, \citenamefont {Babla}, \citenamefont {Le}, \citenamefont {Gang}, \citenamefont {J{\"a}ck}, \citenamefont {Gyenis}, \citenamefont {Yao}, \citenamefont {Cava}, \citenamefont {de~Leon},\ and\ \citenamefont {Houck}}]{Place2021}%
  \BibitemOpen
  \bibfield  {author} {\bibinfo {author} {\bibfnamefont {A.~P.~M.}\ \bibnamefont {Place}}, \bibinfo {author} {\bibfnamefont {L.~V.~H.}\ \bibnamefont {Rodgers}}, \bibinfo {author} {\bibfnamefont {P.}~\bibnamefont {Mundada}}, \bibinfo {author} {\bibfnamefont {B.~M.}\ \bibnamefont {Smitham}}, \bibinfo {author} {\bibfnamefont {M.}~\bibnamefont {Fitzpatrick}}, \bibinfo {author} {\bibfnamefont {Z.}~\bibnamefont {Leng}}, \bibinfo {author} {\bibfnamefont {A.}~\bibnamefont {Premkumar}}, \bibinfo {author} {\bibfnamefont {J.}~\bibnamefont {Bryon}}, \bibinfo {author} {\bibfnamefont {A.}~\bibnamefont {Vrajitoarea}}, \bibinfo {author} {\bibfnamefont {S.}~\bibnamefont {Sussman}}, \bibinfo {author} {\bibfnamefont {G.}~\bibnamefont {Cheng}}, \bibinfo {author} {\bibfnamefont {T.}~\bibnamefont {Madhavan}}, \bibinfo {author} {\bibfnamefont {H.~K.}\ \bibnamefont {Babla}}, \bibinfo {author} {\bibfnamefont {X.~H.}\ \bibnamefont {Le}}, \bibinfo {author} {\bibfnamefont {Y.}~\bibnamefont {Gang}}, \bibinfo {author} {\bibfnamefont
  {B.}~\bibnamefont {J{\"a}ck}}, \bibinfo {author} {\bibfnamefont {A.}~\bibnamefont {Gyenis}}, \bibinfo {author} {\bibfnamefont {N.}~\bibnamefont {Yao}}, \bibinfo {author} {\bibfnamefont {R.~J.}\ \bibnamefont {Cava}}, \bibinfo {author} {\bibfnamefont {N.~P.}\ \bibnamefont {de~Leon}},\ and\ \bibinfo {author} {\bibfnamefont {A.~A.}\ \bibnamefont {Houck}},\ }\bibfield  {title} {\bibinfo {title} {New material platform for superconducting transmon qubits with coherence times exceeding 0.3 milliseconds},\ }\href {https://doi.org/10.1038/s41467-021-22030-5} {\bibfield  {journal} {\bibinfo  {journal} {Nat. Commun.}\ }\textbf {\bibinfo {volume} {12}},\ \bibinfo {pages} {1779} (\bibinfo {year} {2021})}\BibitemShut {NoStop}%
\bibitem [{\citenamefont {Wang}\ \emph {et~al.}(2022)\citenamefont {Wang}, \citenamefont {Li}, \citenamefont {Xu}, \citenamefont {Li}, \citenamefont {Wang}, \citenamefont {Yang}, \citenamefont {Mi}, \citenamefont {Liang}, \citenamefont {Su}, \citenamefont {Yang}, \citenamefont {Wang}, \citenamefont {Wang}, \citenamefont {Li}, \citenamefont {Chen}, \citenamefont {Li}, \citenamefont {Linghu}, \citenamefont {Han}, \citenamefont {Zhang}, \citenamefont {Feng}, \citenamefont {Song}, \citenamefont {Ma}, \citenamefont {Zhang}, \citenamefont {Wang}, \citenamefont {Zhao}, \citenamefont {Liu}, \citenamefont {Xue}, \citenamefont {Jin},\ and\ \citenamefont {Yu}}]{Wang2022}%
  \BibitemOpen
  \bibfield  {author} {\bibinfo {author} {\bibfnamefont {C.}~\bibnamefont {Wang}}, \bibinfo {author} {\bibfnamefont {X.}~\bibnamefont {Li}}, \bibinfo {author} {\bibfnamefont {H.}~\bibnamefont {Xu}}, \bibinfo {author} {\bibfnamefont {Z.}~\bibnamefont {Li}}, \bibinfo {author} {\bibfnamefont {J.}~\bibnamefont {Wang}}, \bibinfo {author} {\bibfnamefont {Z.}~\bibnamefont {Yang}}, \bibinfo {author} {\bibfnamefont {Z.}~\bibnamefont {Mi}}, \bibinfo {author} {\bibfnamefont {X.}~\bibnamefont {Liang}}, \bibinfo {author} {\bibfnamefont {T.}~\bibnamefont {Su}}, \bibinfo {author} {\bibfnamefont {C.}~\bibnamefont {Yang}}, \bibinfo {author} {\bibfnamefont {G.}~\bibnamefont {Wang}}, \bibinfo {author} {\bibfnamefont {W.}~\bibnamefont {Wang}}, \bibinfo {author} {\bibfnamefont {Y.}~\bibnamefont {Li}}, \bibinfo {author} {\bibfnamefont {M.}~\bibnamefont {Chen}}, \bibinfo {author} {\bibfnamefont {C.}~\bibnamefont {Li}}, \bibinfo {author} {\bibfnamefont {K.}~\bibnamefont {Linghu}}, \bibinfo {author} {\bibfnamefont {J.}~\bibnamefont
  {Han}}, \bibinfo {author} {\bibfnamefont {Y.}~\bibnamefont {Zhang}}, \bibinfo {author} {\bibfnamefont {Y.}~\bibnamefont {Feng}}, \bibinfo {author} {\bibfnamefont {Y.}~\bibnamefont {Song}}, \bibinfo {author} {\bibfnamefont {T.}~\bibnamefont {Ma}}, \bibinfo {author} {\bibfnamefont {J.}~\bibnamefont {Zhang}}, \bibinfo {author} {\bibfnamefont {R.}~\bibnamefont {Wang}}, \bibinfo {author} {\bibfnamefont {P.}~\bibnamefont {Zhao}}, \bibinfo {author} {\bibfnamefont {W.}~\bibnamefont {Liu}}, \bibinfo {author} {\bibfnamefont {G.}~\bibnamefont {Xue}}, \bibinfo {author} {\bibfnamefont {Y.}~\bibnamefont {Jin}},\ and\ \bibinfo {author} {\bibfnamefont {H.}~\bibnamefont {Yu}},\ }\bibfield  {title} {\bibinfo {title} {{Towards practical quantum computers: transmon qubit with a lifetime approaching 0.5 milliseconds}},\ }\href {https://doi.org/10.1038/s41534-021-00510-2} {\bibfield  {journal} {\bibinfo  {journal} {npj Quantum Inf.}\ }\textbf {\bibinfo {volume} {8}},\ \bibinfo {pages} {3} (\bibinfo {year} {2022})}\BibitemShut
  {NoStop}%
\bibitem [{\citenamefont {Bal}\ \emph {et~al.}(2024)\citenamefont {Bal}, \citenamefont {Murthy}, \citenamefont {Zhu}, \citenamefont {Crisa}, \citenamefont {You}, \citenamefont {Huang}, \citenamefont {Roy}, \citenamefont {Lee}, \citenamefont {Zanten}, \citenamefont {Pilipenko}, \citenamefont {Nekrashevich}, \citenamefont {Lunin}, \citenamefont {Bafia}, \citenamefont {Krasnikova}, \citenamefont {Kopas}, \citenamefont {Lachman}, \citenamefont {Miller}, \citenamefont {Mutus}, \citenamefont {Reagor}, \citenamefont {Cansizoglu}, \citenamefont {Marshall}, \citenamefont {Pappas}, \citenamefont {Vu}, \citenamefont {Yadavalli}, \citenamefont {Oh}, \citenamefont {Zhou}, \citenamefont {Kramer}, \citenamefont {Lecocq}, \citenamefont {Goronzy}, \citenamefont {Torres-Castanedo}, \citenamefont {Pritchard}, \citenamefont {Dravid}, \citenamefont {Rondinelli}, \citenamefont {Bedzyk}, \citenamefont {Hersam}, \citenamefont {Zasadzinski}, \citenamefont {Koch}, \citenamefont {Sauls}, \citenamefont {Romanenko},\ and\ \citenamefont
  {Grassellino}}]{bal_systematic_2024}%
  \BibitemOpen
  \bibfield  {author} {\bibinfo {author} {\bibfnamefont {M.}~\bibnamefont {Bal}}, \bibinfo {author} {\bibfnamefont {A.~A.}\ \bibnamefont {Murthy}}, \bibinfo {author} {\bibfnamefont {S.}~\bibnamefont {Zhu}}, \bibinfo {author} {\bibfnamefont {F.}~\bibnamefont {Crisa}}, \bibinfo {author} {\bibfnamefont {X.}~\bibnamefont {You}}, \bibinfo {author} {\bibfnamefont {Z.}~\bibnamefont {Huang}}, \bibinfo {author} {\bibfnamefont {T.}~\bibnamefont {Roy}}, \bibinfo {author} {\bibfnamefont {J.}~\bibnamefont {Lee}}, \bibinfo {author} {\bibfnamefont {D.~v.}\ \bibnamefont {Zanten}}, \bibinfo {author} {\bibfnamefont {R.}~\bibnamefont {Pilipenko}}, \bibinfo {author} {\bibfnamefont {I.}~\bibnamefont {Nekrashevich}}, \bibinfo {author} {\bibfnamefont {A.}~\bibnamefont {Lunin}}, \bibinfo {author} {\bibfnamefont {D.}~\bibnamefont {Bafia}}, \bibinfo {author} {\bibfnamefont {Y.}~\bibnamefont {Krasnikova}}, \bibinfo {author} {\bibfnamefont {C.~J.}\ \bibnamefont {Kopas}}, \bibinfo {author} {\bibfnamefont {E.~O.}\ \bibnamefont {Lachman}},
  \bibinfo {author} {\bibfnamefont {D.}~\bibnamefont {Miller}}, \bibinfo {author} {\bibfnamefont {J.~Y.}\ \bibnamefont {Mutus}}, \bibinfo {author} {\bibfnamefont {M.~J.}\ \bibnamefont {Reagor}}, \bibinfo {author} {\bibfnamefont {H.}~\bibnamefont {Cansizoglu}}, \bibinfo {author} {\bibfnamefont {J.}~\bibnamefont {Marshall}}, \bibinfo {author} {\bibfnamefont {D.~P.}\ \bibnamefont {Pappas}}, \bibinfo {author} {\bibfnamefont {K.}~\bibnamefont {Vu}}, \bibinfo {author} {\bibfnamefont {K.}~\bibnamefont {Yadavalli}}, \bibinfo {author} {\bibfnamefont {J.-S.}\ \bibnamefont {Oh}}, \bibinfo {author} {\bibfnamefont {L.}~\bibnamefont {Zhou}}, \bibinfo {author} {\bibfnamefont {M.~J.}\ \bibnamefont {Kramer}}, \bibinfo {author} {\bibfnamefont {F.}~\bibnamefont {Lecocq}}, \bibinfo {author} {\bibfnamefont {D.~P.}\ \bibnamefont {Goronzy}}, \bibinfo {author} {\bibfnamefont {C.~G.}\ \bibnamefont {Torres-Castanedo}}, \bibinfo {author} {\bibfnamefont {P.~G.}\ \bibnamefont {Pritchard}}, \bibinfo {author} {\bibfnamefont {V.~P.}\
  \bibnamefont {Dravid}}, \bibinfo {author} {\bibfnamefont {J.~M.}\ \bibnamefont {Rondinelli}}, \bibinfo {author} {\bibfnamefont {M.~J.}\ \bibnamefont {Bedzyk}}, \bibinfo {author} {\bibfnamefont {M.~C.}\ \bibnamefont {Hersam}}, \bibinfo {author} {\bibfnamefont {J.}~\bibnamefont {Zasadzinski}}, \bibinfo {author} {\bibfnamefont {J.}~\bibnamefont {Koch}}, \bibinfo {author} {\bibfnamefont {J.~A.}\ \bibnamefont {Sauls}}, \bibinfo {author} {\bibfnamefont {A.}~\bibnamefont {Romanenko}},\ and\ \bibinfo {author} {\bibfnamefont {A.}~\bibnamefont {Grassellino}},\ }\bibfield  {title} {\bibinfo {title} {Systematic improvements in transmon qubit coherence enabled by niobium surface encapsulation},\ }\href {https://doi.org/10.1038/s41534-024-00840-x} {\bibfield  {journal} {\bibinfo  {journal} {npj Quantum Inf.}\ }\textbf {\bibinfo {volume} {10}},\ \bibinfo {pages} {1} (\bibinfo {year} {2024})}\BibitemShut {NoStop}%
\bibitem [{\citenamefont {Kalashnikov}\ \emph {et~al.}(2020)\citenamefont {Kalashnikov}, \citenamefont {Hsieh}, \citenamefont {Zhang}, \citenamefont {Lu}, \citenamefont {Kamenov}, \citenamefont {Di~Paolo}, \citenamefont {Blais}, \citenamefont {Gershenson},\ and\ \citenamefont {Bell}}]{kalashnikov2020bifluxon}%
  \BibitemOpen
  \bibfield  {author} {\bibinfo {author} {\bibfnamefont {K.}~\bibnamefont {Kalashnikov}}, \bibinfo {author} {\bibfnamefont {W.~T.}\ \bibnamefont {Hsieh}}, \bibinfo {author} {\bibfnamefont {W.}~\bibnamefont {Zhang}}, \bibinfo {author} {\bibfnamefont {W.-S.}\ \bibnamefont {Lu}}, \bibinfo {author} {\bibfnamefont {P.}~\bibnamefont {Kamenov}}, \bibinfo {author} {\bibfnamefont {A.}~\bibnamefont {Di~Paolo}}, \bibinfo {author} {\bibfnamefont {A.}~\bibnamefont {Blais}}, \bibinfo {author} {\bibfnamefont {M.~E.}\ \bibnamefont {Gershenson}},\ and\ \bibinfo {author} {\bibfnamefont {M.}~\bibnamefont {Bell}},\ }\bibfield  {title} {\bibinfo {title} {Bifluxon: Fluxon-parity-protected superconducting qubit},\ }\href {https://doi.org/10.1103/PRXQuantum.1.010307} {\bibfield  {journal} {\bibinfo  {journal} {PRX Quantum}\ }\textbf {\bibinfo {volume} {1}},\ \bibinfo {pages} {010307} (\bibinfo {year} {2020})}\BibitemShut {NoStop}%
\bibitem [{\citenamefont {Gyenis}\ \emph {et~al.}(2021{\natexlab{a}})\citenamefont {Gyenis}, \citenamefont {Mundada}, \citenamefont {Di~Paolo}, \citenamefont {Hazard}, \citenamefont {You}, \citenamefont {Schuster}, \citenamefont {Koch}, \citenamefont {Blais},\ and\ \citenamefont {Houck}}]{Gyenis2021}%
  \BibitemOpen
  \bibfield  {author} {\bibinfo {author} {\bibfnamefont {A.}~\bibnamefont {Gyenis}}, \bibinfo {author} {\bibfnamefont {P.~S.}\ \bibnamefont {Mundada}}, \bibinfo {author} {\bibfnamefont {A.}~\bibnamefont {Di~Paolo}}, \bibinfo {author} {\bibfnamefont {T.~M.}\ \bibnamefont {Hazard}}, \bibinfo {author} {\bibfnamefont {X.}~\bibnamefont {You}}, \bibinfo {author} {\bibfnamefont {D.~I.}\ \bibnamefont {Schuster}}, \bibinfo {author} {\bibfnamefont {J.}~\bibnamefont {Koch}}, \bibinfo {author} {\bibfnamefont {A.}~\bibnamefont {Blais}},\ and\ \bibinfo {author} {\bibfnamefont {A.~A.}\ \bibnamefont {Houck}},\ }\bibfield  {title} {\bibinfo {title} {Experimental realization of a protected superconducting circuit derived from the $0$--$\ensuremath{\pi}$ qubit},\ }\href {https://doi.org/10.1103/PRXQuantum.2.010339} {\bibfield  {journal} {\bibinfo  {journal} {PRX Quantum}\ }\textbf {\bibinfo {volume} {2}},\ \bibinfo {pages} {010339} (\bibinfo {year} {2021}{\natexlab{a}})}\BibitemShut {NoStop}%
\bibitem [{\citenamefont {Gyenis}\ \emph {et~al.}(2021{\natexlab{b}})\citenamefont {Gyenis}, \citenamefont {Di~Paolo}, \citenamefont {Koch}, \citenamefont {Blais}, \citenamefont {Houck},\ and\ \citenamefont {Schuster}}]{GyenisReview2021}%
  \BibitemOpen
  \bibfield  {author} {\bibinfo {author} {\bibfnamefont {A.}~\bibnamefont {Gyenis}}, \bibinfo {author} {\bibfnamefont {A.}~\bibnamefont {Di~Paolo}}, \bibinfo {author} {\bibfnamefont {J.}~\bibnamefont {Koch}}, \bibinfo {author} {\bibfnamefont {A.}~\bibnamefont {Blais}}, \bibinfo {author} {\bibfnamefont {A.~A.}\ \bibnamefont {Houck}},\ and\ \bibinfo {author} {\bibfnamefont {D.~I.}\ \bibnamefont {Schuster}},\ }\bibfield  {title} {\bibinfo {title} {Moving beyond the transmon: Noise-protected superconducting quantum circuits},\ }\href {https://doi.org/10.1103/PRXQuantum.2.030101} {\bibfield  {journal} {\bibinfo  {journal} {PRX Quantum}\ }\textbf {\bibinfo {volume} {2}},\ \bibinfo {pages} {030101} (\bibinfo {year} {2021}{\natexlab{b}})}\BibitemShut {NoStop}%
\bibitem [{\citenamefont {Kitaev}(2003)}]{Kitaev2003}%
  \BibitemOpen
  \bibfield  {author} {\bibinfo {author} {\bibfnamefont {A.}~\bibnamefont {Kitaev}},\ }\bibfield  {title} {\bibinfo {title} {Fault-tolerant quantum computation by anyons},\ }\href {https://doi.org/https://doi.org/10.1016/S0003-4916(02)00018-0} {\bibfield  {journal} {\bibinfo  {journal} {Ann. Phys.}\ }\textbf {\bibinfo {volume} {303}},\ \bibinfo {pages} {2} (\bibinfo {year} {2003})}\BibitemShut {NoStop}%
\bibitem [{\citenamefont {Larsen}\ \emph {et~al.}(2020)\citenamefont {Larsen}, \citenamefont {Gershenson}, \citenamefont {Casparis}, \citenamefont {Kringh\o{}j}, \citenamefont {Pearson}, \citenamefont {McNeil}, \citenamefont {Kuemmeth}, \citenamefont {Krogstrup}, \citenamefont {Petersson},\ and\ \citenamefont {Marcus}}]{PhysRevLett.125.056801}%
  \BibitemOpen
  \bibfield  {author} {\bibinfo {author} {\bibfnamefont {T.~W.}\ \bibnamefont {Larsen}}, \bibinfo {author} {\bibfnamefont {M.~E.}\ \bibnamefont {Gershenson}}, \bibinfo {author} {\bibfnamefont {L.}~\bibnamefont {Casparis}}, \bibinfo {author} {\bibfnamefont {A.}~\bibnamefont {Kringh\o{}j}}, \bibinfo {author} {\bibfnamefont {N.~J.}\ \bibnamefont {Pearson}}, \bibinfo {author} {\bibfnamefont {R.~P.~G.}\ \bibnamefont {McNeil}}, \bibinfo {author} {\bibfnamefont {F.}~\bibnamefont {Kuemmeth}}, \bibinfo {author} {\bibfnamefont {P.}~\bibnamefont {Krogstrup}}, \bibinfo {author} {\bibfnamefont {K.~D.}\ \bibnamefont {Petersson}},\ and\ \bibinfo {author} {\bibfnamefont {C.~M.}\ \bibnamefont {Marcus}},\ }\bibfield  {title} {\bibinfo {title} {Parity-protected superconductor-semiconductor qubit},\ }\href {https://doi.org/10.1103/PhysRevLett.125.056801} {\bibfield  {journal} {\bibinfo  {journal} {Phys. Rev. Lett.}\ }\textbf {\bibinfo {volume} {125}},\ \bibinfo {pages} {056801} (\bibinfo {year} {2020})}\BibitemShut {NoStop}%
\bibitem [{\citenamefont {Schrade}\ \emph {et~al.}(2022)\citenamefont {Schrade}, \citenamefont {Marcus},\ and\ \citenamefont {Gyenis}}]{Schrade2022}%
  \BibitemOpen
  \bibfield  {author} {\bibinfo {author} {\bibfnamefont {C.}~\bibnamefont {Schrade}}, \bibinfo {author} {\bibfnamefont {C.~M.}\ \bibnamefont {Marcus}},\ and\ \bibinfo {author} {\bibfnamefont {A.}~\bibnamefont {Gyenis}},\ }\bibfield  {title} {\bibinfo {title} {Protected hybrid superconducting qubit in an array of gate-tunable josephson interferometers},\ }\href {https://doi.org/10.1103/PRXQuantum.3.030303} {\bibfield  {journal} {\bibinfo  {journal} {PRX Quantum}\ }\textbf {\bibinfo {volume} {3}},\ \bibinfo {pages} {030303} (\bibinfo {year} {2022})}\BibitemShut {NoStop}%
\bibitem [{\citenamefont {Kou}\ \emph {et~al.}(2017)\citenamefont {Kou}, \citenamefont {Smith}, \citenamefont {Vool}, \citenamefont {Brierley}, \citenamefont {Meier}, \citenamefont {Frunzio}, \citenamefont {Girvin}, \citenamefont {Glazman},\ and\ \citenamefont {Devoret}}]{kou2017}%
  \BibitemOpen
  \bibfield  {author} {\bibinfo {author} {\bibfnamefont {A.}~\bibnamefont {Kou}}, \bibinfo {author} {\bibfnamefont {W.~C.}\ \bibnamefont {Smith}}, \bibinfo {author} {\bibfnamefont {U.}~\bibnamefont {Vool}}, \bibinfo {author} {\bibfnamefont {R.~T.}\ \bibnamefont {Brierley}}, \bibinfo {author} {\bibfnamefont {H.}~\bibnamefont {Meier}}, \bibinfo {author} {\bibfnamefont {L.}~\bibnamefont {Frunzio}}, \bibinfo {author} {\bibfnamefont {S.~M.}\ \bibnamefont {Girvin}}, \bibinfo {author} {\bibfnamefont {L.~I.}\ \bibnamefont {Glazman}},\ and\ \bibinfo {author} {\bibfnamefont {M.~H.}\ \bibnamefont {Devoret}},\ }\bibfield  {title} {\bibinfo {title} {Fluxonium-based artificial molecule with a tunable magnetic moment},\ }\href {https://doi.org/10.1103/PhysRevX.7.031037} {\bibfield  {journal} {\bibinfo  {journal} {Phys. Rev. X}\ }\textbf {\bibinfo {volume} {7}},\ \bibinfo {pages} {031037} (\bibinfo {year} {2017})}\BibitemShut {NoStop}%
\bibitem [{\citenamefont {You}(2021)}]{YouThesis}%
  \BibitemOpen
  \bibfield  {author} {\bibinfo {author} {\bibfnamefont {X.}~\bibnamefont {You}},\ }\emph {\bibinfo {title} {Noise-Protected Superconducting Quantum Circuits}},\ \href@noop {} {Ph.D. thesis},\ \bibinfo  {school} {Northwestern Univ.} (\bibinfo {year} {2021})\BibitemShut {NoStop}%
\bibitem [{\citenamefont {Thibodeau}\ \emph {et~al.}(2024)\citenamefont {Thibodeau}, \citenamefont {Kou},\ and\ \citenamefont {Clark}}]{thibodeau2024floquet}%
  \BibitemOpen
  \bibfield  {author} {\bibinfo {author} {\bibfnamefont {M.}~\bibnamefont {Thibodeau}}, \bibinfo {author} {\bibfnamefont {A.}~\bibnamefont {Kou}},\ and\ \bibinfo {author} {\bibfnamefont {B.~K.}\ \bibnamefont {Clark}},\ }\bibfield  {title} {\bibinfo {title} {The floquet fluxonium molecule: Driving down dephasing in coupled superconducting qubits},\ }\href {https://doi.org/10.1103/PRXQuantum.5.040314} {\bibfield  {journal} {\bibinfo  {journal} {PRX Quantum}\ }\textbf {\bibinfo {volume} {5}},\ \bibinfo {pages} {040314} (\bibinfo {year} {2024})}\BibitemShut {NoStop}%
\bibitem [{\citenamefont {Brooks}\ \emph {et~al.}(2013)\citenamefont {Brooks}, \citenamefont {Kitaev},\ and\ \citenamefont {Preskill}}]{PhysRevA.87.052306}%
  \BibitemOpen
  \bibfield  {author} {\bibinfo {author} {\bibfnamefont {P.}~\bibnamefont {Brooks}}, \bibinfo {author} {\bibfnamefont {A.}~\bibnamefont {Kitaev}},\ and\ \bibinfo {author} {\bibfnamefont {J.}~\bibnamefont {Preskill}},\ }\bibfield  {title} {\bibinfo {title} {Protected gates for superconducting qubits},\ }\href {https://doi.org/10.1103/PhysRevA.87.052306} {\bibfield  {journal} {\bibinfo  {journal} {Phys. Rev. A}\ }\textbf {\bibinfo {volume} {87}},\ \bibinfo {pages} {052306} (\bibinfo {year} {2013})}\BibitemShut {NoStop}%
\bibitem [{\citenamefont {Groszkowski}\ \emph {et~al.}(2018)\citenamefont {Groszkowski}, \citenamefont {Paolo}, \citenamefont {Grimsmo}, \citenamefont {Blais}, \citenamefont {Schuster}, \citenamefont {Houck},\ and\ \citenamefont {Koch}}]{Groszkowski2018}%
  \BibitemOpen
  \bibfield  {author} {\bibinfo {author} {\bibfnamefont {P.}~\bibnamefont {Groszkowski}}, \bibinfo {author} {\bibfnamefont {A.~D.}\ \bibnamefont {Paolo}}, \bibinfo {author} {\bibfnamefont {A.~L.}\ \bibnamefont {Grimsmo}}, \bibinfo {author} {\bibfnamefont {A.}~\bibnamefont {Blais}}, \bibinfo {author} {\bibfnamefont {D.~I.}\ \bibnamefont {Schuster}}, \bibinfo {author} {\bibfnamefont {A.~A.}\ \bibnamefont {Houck}},\ and\ \bibinfo {author} {\bibfnamefont {J.}~\bibnamefont {Koch}},\ }\bibfield  {title} {\bibinfo {title} {Coherence properties of the $0$--$\ensuremath{\pi}$ qubit},\ }\href {https://doi.org/10.1088/1367-2630/aab7cd} {\bibfield  {journal} {\bibinfo  {journal} {New J. Phys.}\ }\textbf {\bibinfo {volume} {20}},\ \bibinfo {pages} {043053} (\bibinfo {year} {2018})}\BibitemShut {NoStop}%
\bibitem [{\citenamefont {Rymarz}\ and\ \citenamefont {DiVincenzo}(2023)}]{PhysRevX.13.021017}%
  \BibitemOpen
  \bibfield  {author} {\bibinfo {author} {\bibfnamefont {M.}~\bibnamefont {Rymarz}}\ and\ \bibinfo {author} {\bibfnamefont {D.~P.}\ \bibnamefont {DiVincenzo}},\ }\bibfield  {title} {\bibinfo {title} {Consistent quantization of nearly singular superconducting circuits},\ }\href {https://doi.org/10.1103/PhysRevX.13.021017} {\bibfield  {journal} {\bibinfo  {journal} {Phys. Rev. X}\ }\textbf {\bibinfo {volume} {13}},\ \bibinfo {pages} {021017} (\bibinfo {year} {2023})}\BibitemShut {NoStop}%
\bibitem [{\citenamefont {Koch}\ \emph {et~al.}(2009)\citenamefont {Koch}, \citenamefont {Manucharyan}, \citenamefont {Devoret},\ and\ \citenamefont {Glazman}}]{Koch2009}%
  \BibitemOpen
  \bibfield  {author} {\bibinfo {author} {\bibfnamefont {J.}~\bibnamefont {Koch}}, \bibinfo {author} {\bibfnamefont {V.}~\bibnamefont {Manucharyan}}, \bibinfo {author} {\bibfnamefont {M.~H.}\ \bibnamefont {Devoret}},\ and\ \bibinfo {author} {\bibfnamefont {L.~I.}\ \bibnamefont {Glazman}},\ }\bibfield  {title} {\bibinfo {title} {{Charging Effects in the Inductively Shunted Josephson Junction}},\ }\href {https://doi.org/10.1103/PhysRevLett.103.217004} {\bibfield  {journal} {\bibinfo  {journal} {Phys. Rev. Lett.}\ }\textbf {\bibinfo {volume} {103}},\ \bibinfo {pages} {217004} (\bibinfo {year} {2009})}\BibitemShut {NoStop}%
\bibitem [{\citenamefont {Groszkowski}\ and\ \citenamefont {Koch}(2021)}]{Groszkowski2021}%
  \BibitemOpen
  \bibfield  {author} {\bibinfo {author} {\bibfnamefont {P.}~\bibnamefont {Groszkowski}}\ and\ \bibinfo {author} {\bibfnamefont {J.}~\bibnamefont {Koch}},\ }\bibfield  {title} {\bibinfo {title} {Scqubits: a {P}ython package for superconducting qubits},\ }\href {https://doi.org/10.22331/q-2021-11-17-583} {\bibfield  {journal} {\bibinfo  {journal} {{Quantum}}\ }\textbf {\bibinfo {volume} {5}},\ \bibinfo {pages} {583} (\bibinfo {year} {2021})}\BibitemShut {NoStop}%
\bibitem [{\citenamefont {Chitta}\ \emph {et~al.}(2022)\citenamefont {Chitta}, \citenamefont {Zhao}, \citenamefont {Huang}, \citenamefont {Mondragon-Shem},\ and\ \citenamefont {Koch}}]{Chitta2022}%
  \BibitemOpen
  \bibfield  {author} {\bibinfo {author} {\bibfnamefont {S.~P.}\ \bibnamefont {Chitta}}, \bibinfo {author} {\bibfnamefont {T.}~\bibnamefont {Zhao}}, \bibinfo {author} {\bibfnamefont {Z.}~\bibnamefont {Huang}}, \bibinfo {author} {\bibfnamefont {I.}~\bibnamefont {Mondragon-Shem}},\ and\ \bibinfo {author} {\bibfnamefont {J.}~\bibnamefont {Koch}},\ }\bibfield  {title} {\bibinfo {title} {Computer-aided quantization and numerical analysis of superconducting circuits},\ }\href {https://doi.org/10.1088/1367-2630/ac94f2} {\bibfield  {journal} {\bibinfo  {journal} {New J. Phys.}\ }\textbf {\bibinfo {volume} {24}},\ \bibinfo {pages} {103020} (\bibinfo {year} {2022})}\BibitemShut {NoStop}%
\bibitem [{\citenamefont {Kubica}\ \emph {et~al.}(2023)\citenamefont {Kubica}, \citenamefont {Haim}, \citenamefont {Vaknin}, \citenamefont {Levine}, \citenamefont {Brand{\~{a}}o},\ and\ \citenamefont {Retzker}}]{Kubica2023}%
  \BibitemOpen
  \bibfield  {author} {\bibinfo {author} {\bibfnamefont {A.}~\bibnamefont {Kubica}}, \bibinfo {author} {\bibfnamefont {A.}~\bibnamefont {Haim}}, \bibinfo {author} {\bibfnamefont {Y.}~\bibnamefont {Vaknin}}, \bibinfo {author} {\bibfnamefont {H.}~\bibnamefont {Levine}}, \bibinfo {author} {\bibfnamefont {F.}~\bibnamefont {Brand{\~{a}}o}},\ and\ \bibinfo {author} {\bibfnamefont {A.}~\bibnamefont {Retzker}},\ }\bibfield  {title} {\bibinfo {title} {{Erasure Qubits: Overcoming the T1 Limit in Superconducting Circuits}},\ }\href {https://doi.org/10.1103/PhysRevX.13.041022} {\bibfield  {journal} {\bibinfo  {journal} {Phys. Rev. X}\ }\textbf {\bibinfo {volume} {13}},\ \bibinfo {pages} {41022} (\bibinfo {year} {2023})}\BibitemShut {NoStop}%
\bibitem [{\citenamefont {Chou}\ \emph {et~al.}(2023)\citenamefont {Chou}, \citenamefont {Shemma}, \citenamefont {McCarrick}, \citenamefont {Chien}, \citenamefont {Teoh}, \citenamefont {Winkel}, \citenamefont {Anderson}, \citenamefont {Chen}, \citenamefont {Curtis}, \citenamefont {de~Graaf}, \citenamefont {Garmon}, \citenamefont {Gudlewski}, \citenamefont {Kalfus}, \citenamefont {Keen}, \citenamefont {Khedkar}, \citenamefont {Lei}, \citenamefont {Liu}, \citenamefont {Lu}, \citenamefont {Lu}, \citenamefont {Maiti}, \citenamefont {Mastalli-Kelly}, \citenamefont {Mehta}, \citenamefont {Mundhada}, \citenamefont {Narla}, \citenamefont {Noh}, \citenamefont {Tsunoda}, \citenamefont {Xue}, \citenamefont {Yuan}, \citenamefont {Frunzio}, \citenamefont {Aumentado}, \citenamefont {Puri}, \citenamefont {Girvin}, \citenamefont {Moseley},\ and\ \citenamefont {Schoelkopf}}]{Chou2023}%
  \BibitemOpen
  \bibfield  {author} {\bibinfo {author} {\bibfnamefont {K.~S.}\ \bibnamefont {Chou}}, \bibinfo {author} {\bibfnamefont {T.}~\bibnamefont {Shemma}}, \bibinfo {author} {\bibfnamefont {H.}~\bibnamefont {McCarrick}}, \bibinfo {author} {\bibfnamefont {T.-C.}\ \bibnamefont {Chien}}, \bibinfo {author} {\bibfnamefont {J.~D.}\ \bibnamefont {Teoh}}, \bibinfo {author} {\bibfnamefont {P.}~\bibnamefont {Winkel}}, \bibinfo {author} {\bibfnamefont {A.}~\bibnamefont {Anderson}}, \bibinfo {author} {\bibfnamefont {J.}~\bibnamefont {Chen}}, \bibinfo {author} {\bibfnamefont {J.}~\bibnamefont {Curtis}}, \bibinfo {author} {\bibfnamefont {S.~J.}\ \bibnamefont {de~Graaf}}, \bibinfo {author} {\bibfnamefont {J.~W.~O.}\ \bibnamefont {Garmon}}, \bibinfo {author} {\bibfnamefont {B.}~\bibnamefont {Gudlewski}}, \bibinfo {author} {\bibfnamefont {W.~D.}\ \bibnamefont {Kalfus}}, \bibinfo {author} {\bibfnamefont {T.}~\bibnamefont {Keen}}, \bibinfo {author} {\bibfnamefont {N.}~\bibnamefont {Khedkar}}, \bibinfo {author} {\bibfnamefont {C.~U.}\
  \bibnamefont {Lei}}, \bibinfo {author} {\bibfnamefont {G.}~\bibnamefont {Liu}}, \bibinfo {author} {\bibfnamefont {P.}~\bibnamefont {Lu}}, \bibinfo {author} {\bibfnamefont {Y.}~\bibnamefont {Lu}}, \bibinfo {author} {\bibfnamefont {A.}~\bibnamefont {Maiti}}, \bibinfo {author} {\bibfnamefont {L.}~\bibnamefont {Mastalli-Kelly}}, \bibinfo {author} {\bibfnamefont {N.}~\bibnamefont {Mehta}}, \bibinfo {author} {\bibfnamefont {S.~O.}\ \bibnamefont {Mundhada}}, \bibinfo {author} {\bibfnamefont {A.}~\bibnamefont {Narla}}, \bibinfo {author} {\bibfnamefont {T.}~\bibnamefont {Noh}}, \bibinfo {author} {\bibfnamefont {T.}~\bibnamefont {Tsunoda}}, \bibinfo {author} {\bibfnamefont {S.~H.}\ \bibnamefont {Xue}}, \bibinfo {author} {\bibfnamefont {J.~O.}\ \bibnamefont {Yuan}}, \bibinfo {author} {\bibfnamefont {L.}~\bibnamefont {Frunzio}}, \bibinfo {author} {\bibfnamefont {J.}~\bibnamefont {Aumentado}}, \bibinfo {author} {\bibfnamefont {S.}~\bibnamefont {Puri}}, \bibinfo {author} {\bibfnamefont {S.~M.}\ \bibnamefont {Girvin}},
  \bibinfo {author} {\bibfnamefont {S.~H.}\ \bibnamefont {Moseley}},\ and\ \bibinfo {author} {\bibfnamefont {R.~J.}\ \bibnamefont {Schoelkopf}},\ }\bibfield  {title} {\bibinfo {title} {{Demonstrating a superconducting dual-rail cavity qubit with erasure-detected logical measurements}},\ }\href {https://arxiv.org/abs/2307.03169} {\bibfield  {journal} {\bibinfo  {journal} {arXiv:2307.03169}\ } (\bibinfo {year} {2023})}\BibitemShut {NoStop}%
\bibitem [{\citenamefont {Teoh}\ \emph {et~al.}(2023)\citenamefont {Teoh}, \citenamefont {Winkel}, \citenamefont {Babla}, \citenamefont {Chapman}, \citenamefont {Claes}, \citenamefont {de~Graaf}, \citenamefont {Garmon}, \citenamefont {Kalfus}, \citenamefont {Lu}, \citenamefont {Maiti}, \citenamefont {Sahay}, \citenamefont {Thakur}, \citenamefont {Tsunoda}, \citenamefont {Xue}, \citenamefont {Frunzio}, \citenamefont {Girvin}, \citenamefont {Puri},\ and\ \citenamefont {Schoelkopf}}]{Teoh2023}%
  \BibitemOpen
  \bibfield  {author} {\bibinfo {author} {\bibfnamefont {J.~D.}\ \bibnamefont {Teoh}}, \bibinfo {author} {\bibfnamefont {P.}~\bibnamefont {Winkel}}, \bibinfo {author} {\bibfnamefont {H.~K.}\ \bibnamefont {Babla}}, \bibinfo {author} {\bibfnamefont {B.~J.}\ \bibnamefont {Chapman}}, \bibinfo {author} {\bibfnamefont {J.}~\bibnamefont {Claes}}, \bibinfo {author} {\bibfnamefont {S.~J.}\ \bibnamefont {de~Graaf}}, \bibinfo {author} {\bibfnamefont {J.~W.~O.}\ \bibnamefont {Garmon}}, \bibinfo {author} {\bibfnamefont {W.~D.}\ \bibnamefont {Kalfus}}, \bibinfo {author} {\bibfnamefont {Y.}~\bibnamefont {Lu}}, \bibinfo {author} {\bibfnamefont {A.}~\bibnamefont {Maiti}}, \bibinfo {author} {\bibfnamefont {K.}~\bibnamefont {Sahay}}, \bibinfo {author} {\bibfnamefont {N.}~\bibnamefont {Thakur}}, \bibinfo {author} {\bibfnamefont {T.}~\bibnamefont {Tsunoda}}, \bibinfo {author} {\bibfnamefont {S.~H.}\ \bibnamefont {Xue}}, \bibinfo {author} {\bibfnamefont {L.}~\bibnamefont {Frunzio}}, \bibinfo {author} {\bibfnamefont {S.~M.}\
  \bibnamefont {Girvin}}, \bibinfo {author} {\bibfnamefont {S.}~\bibnamefont {Puri}},\ and\ \bibinfo {author} {\bibfnamefont {R.~J.}\ \bibnamefont {Schoelkopf}},\ }\bibfield  {title} {\bibinfo {title} {{Dual-rail encoding with superconducting cavities}},\ }\href {https://pnas.org/doi/10.1073/pnas.2221736120} {\bibfield  {journal} {\bibinfo  {journal} {Proc. Natl. Acad. Sci.}\ }\textbf {\bibinfo {volume} {120}} (\bibinfo {year} {2023})}\BibitemShut {NoStop}%
\bibitem [{\citenamefont {Levine}\ \emph {et~al.}(2024)\citenamefont {Levine}, \citenamefont {Haim}, \citenamefont {Hung}, \citenamefont {Alidoust}, \citenamefont {Kalaee}, \citenamefont {DeLorenzo}, \citenamefont {Wollack}, \citenamefont {Arrangoiz-Arriola}, \citenamefont {Khalajhedayati}, \citenamefont {Sanil}, \citenamefont {Moradinejad}, \citenamefont {Vaknin}, \citenamefont {Kubica}, \citenamefont {Hover}, \citenamefont {Aghaeimeibodi}, \citenamefont {Alcid}, \citenamefont {Baek}, \citenamefont {Barnett}, \citenamefont {Bawdekar}, \citenamefont {Bienias}, \citenamefont {Carson}, \citenamefont {Chen}, \citenamefont {Chen}, \citenamefont {Chinkezian}, \citenamefont {Chisholm}, \citenamefont {Clifford}, \citenamefont {Cosmic}, \citenamefont {Crisosto}, \citenamefont {Dalzell}, \citenamefont {Davis}, \citenamefont {D'Ewart}, \citenamefont {Diez}, \citenamefont {D'Souza}, \citenamefont {Dumitrescu}, \citenamefont {Elkhouly}, \citenamefont {Fang}, \citenamefont {Fang}, \citenamefont {Flammia}, \citenamefont
  {Fling}, \citenamefont {Garcia}, \citenamefont {Gharzai}, \citenamefont {Gorshkov}, \citenamefont {Gray}, \citenamefont {Grimberg}, \citenamefont {Grimsmo}, \citenamefont {Hann}, \citenamefont {He}, \citenamefont {Heidel}, \citenamefont {Howell}, \citenamefont {Hunt}, \citenamefont {Iverson}, \citenamefont {Jarrige}, \citenamefont {Jiang}, \citenamefont {Jones}, \citenamefont {Karabalin}, \citenamefont {Karalekas}, \citenamefont {Keller}, \citenamefont {Lasi}, \citenamefont {Lee}, \citenamefont {Ly}, \citenamefont {MacCabe}, \citenamefont {Mahuli}, \citenamefont {Marcaud}, \citenamefont {Matheny}, \citenamefont {McArdle}, \citenamefont {McCabe}, \citenamefont {Merton}, \citenamefont {Miles}, \citenamefont {Milsted}, \citenamefont {Mishra}, \citenamefont {Moncelsi}, \citenamefont {Naghiloo}, \citenamefont {Noh}, \citenamefont {Oblepias}, \citenamefont {Ortuno}, \citenamefont {Owens}, \citenamefont {Pagdilao}, \citenamefont {Panduro}, \citenamefont {Paquette}, \citenamefont {Patel}, \citenamefont {Peairs},
  \citenamefont {Perello}, \citenamefont {Peterson}, \citenamefont {Ponte}, \citenamefont {Putterman}, \citenamefont {Refael}, \citenamefont {Reinhold}, \citenamefont {Resnick}, \citenamefont {Reyna}, \citenamefont {Rodriguez}, \citenamefont {Rose}, \citenamefont {Rubin}, \citenamefont {Runyan}, \citenamefont {Ryan}, \citenamefont {Sahmoud}, \citenamefont {Scaffidi}, \citenamefont {Shah}, \citenamefont {Siavoshi}, \citenamefont {Sivarajah}, \citenamefont {Skogland}, \citenamefont {Su}, \citenamefont {Swenson}, \citenamefont {Sylvia}, \citenamefont {Teo}, \citenamefont {Tomada}, \citenamefont {Torlai}, \citenamefont {Wistrom}, \citenamefont {Zhang}, \citenamefont {Zuk}, \citenamefont {Clerk}, \citenamefont {Brand{\~{a}}o}, \citenamefont {Retzker},\ and\ \citenamefont {Painter}}]{Levine2024}%
  \BibitemOpen
  \bibfield  {author} {\bibinfo {author} {\bibfnamefont {H.}~\bibnamefont {Levine}}, \bibinfo {author} {\bibfnamefont {A.}~\bibnamefont {Haim}}, \bibinfo {author} {\bibfnamefont {J.~S.~C.}\ \bibnamefont {Hung}}, \bibinfo {author} {\bibfnamefont {N.}~\bibnamefont {Alidoust}}, \bibinfo {author} {\bibfnamefont {M.}~\bibnamefont {Kalaee}}, \bibinfo {author} {\bibfnamefont {L.}~\bibnamefont {DeLorenzo}}, \bibinfo {author} {\bibfnamefont {E.~A.}\ \bibnamefont {Wollack}}, \bibinfo {author} {\bibfnamefont {P.}~\bibnamefont {Arrangoiz-Arriola}}, \bibinfo {author} {\bibfnamefont {A.}~\bibnamefont {Khalajhedayati}}, \bibinfo {author} {\bibfnamefont {R.}~\bibnamefont {Sanil}}, \bibinfo {author} {\bibfnamefont {H.}~\bibnamefont {Moradinejad}}, \bibinfo {author} {\bibfnamefont {Y.}~\bibnamefont {Vaknin}}, \bibinfo {author} {\bibfnamefont {A.}~\bibnamefont {Kubica}}, \bibinfo {author} {\bibfnamefont {D.}~\bibnamefont {Hover}}, \bibinfo {author} {\bibfnamefont {S.}~\bibnamefont {Aghaeimeibodi}}, \bibinfo {author}
  {\bibfnamefont {J.~A.}\ \bibnamefont {Alcid}}, \bibinfo {author} {\bibfnamefont {C.}~\bibnamefont {Baek}}, \bibinfo {author} {\bibfnamefont {J.}~\bibnamefont {Barnett}}, \bibinfo {author} {\bibfnamefont {K.}~\bibnamefont {Bawdekar}}, \bibinfo {author} {\bibfnamefont {P.}~\bibnamefont {Bienias}}, \bibinfo {author} {\bibfnamefont {H.~A.}\ \bibnamefont {Carson}}, \bibinfo {author} {\bibfnamefont {C.}~\bibnamefont {Chen}}, \bibinfo {author} {\bibfnamefont {L.}~\bibnamefont {Chen}}, \bibinfo {author} {\bibfnamefont {H.}~\bibnamefont {Chinkezian}}, \bibinfo {author} {\bibfnamefont {E.~M.}\ \bibnamefont {Chisholm}}, \bibinfo {author} {\bibfnamefont {A.}~\bibnamefont {Clifford}}, \bibinfo {author} {\bibfnamefont {R.}~\bibnamefont {Cosmic}}, \bibinfo {author} {\bibfnamefont {N.}~\bibnamefont {Crisosto}}, \bibinfo {author} {\bibfnamefont {A.~M.}\ \bibnamefont {Dalzell}}, \bibinfo {author} {\bibfnamefont {E.}~\bibnamefont {Davis}}, \bibinfo {author} {\bibfnamefont {J.~M.}\ \bibnamefont {D'Ewart}}, \bibinfo {author}
  {\bibfnamefont {S.}~\bibnamefont {Diez}}, \bibinfo {author} {\bibfnamefont {N.}~\bibnamefont {D'Souza}}, \bibinfo {author} {\bibfnamefont {P.~T.}\ \bibnamefont {Dumitrescu}}, \bibinfo {author} {\bibfnamefont {E.}~\bibnamefont {Elkhouly}}, \bibinfo {author} {\bibfnamefont {M.~T.}\ \bibnamefont {Fang}}, \bibinfo {author} {\bibfnamefont {Y.}~\bibnamefont {Fang}}, \bibinfo {author} {\bibfnamefont {S.}~\bibnamefont {Flammia}}, \bibinfo {author} {\bibfnamefont {M.~J.}\ \bibnamefont {Fling}}, \bibinfo {author} {\bibfnamefont {G.}~\bibnamefont {Garcia}}, \bibinfo {author} {\bibfnamefont {M.~K.}\ \bibnamefont {Gharzai}}, \bibinfo {author} {\bibfnamefont {A.~V.}\ \bibnamefont {Gorshkov}}, \bibinfo {author} {\bibfnamefont {M.~J.}\ \bibnamefont {Gray}}, \bibinfo {author} {\bibfnamefont {S.}~\bibnamefont {Grimberg}}, \bibinfo {author} {\bibfnamefont {A.~L.}\ \bibnamefont {Grimsmo}}, \bibinfo {author} {\bibfnamefont {C.~T.}\ \bibnamefont {Hann}}, \bibinfo {author} {\bibfnamefont {Y.}~\bibnamefont {He}}, \bibinfo {author}
  {\bibfnamefont {S.}~\bibnamefont {Heidel}}, \bibinfo {author} {\bibfnamefont {S.}~\bibnamefont {Howell}}, \bibinfo {author} {\bibfnamefont {M.}~\bibnamefont {Hunt}}, \bibinfo {author} {\bibfnamefont {J.}~\bibnamefont {Iverson}}, \bibinfo {author} {\bibfnamefont {I.}~\bibnamefont {Jarrige}}, \bibinfo {author} {\bibfnamefont {L.}~\bibnamefont {Jiang}}, \bibinfo {author} {\bibfnamefont {W.~M.}\ \bibnamefont {Jones}}, \bibinfo {author} {\bibfnamefont {R.}~\bibnamefont {Karabalin}}, \bibinfo {author} {\bibfnamefont {P.~J.}\ \bibnamefont {Karalekas}}, \bibinfo {author} {\bibfnamefont {A.~J.}\ \bibnamefont {Keller}}, \bibinfo {author} {\bibfnamefont {D.}~\bibnamefont {Lasi}}, \bibinfo {author} {\bibfnamefont {M.}~\bibnamefont {Lee}}, \bibinfo {author} {\bibfnamefont {V.}~\bibnamefont {Ly}}, \bibinfo {author} {\bibfnamefont {G.}~\bibnamefont {MacCabe}}, \bibinfo {author} {\bibfnamefont {N.}~\bibnamefont {Mahuli}}, \bibinfo {author} {\bibfnamefont {G.}~\bibnamefont {Marcaud}}, \bibinfo {author} {\bibfnamefont
  {M.~H.}\ \bibnamefont {Matheny}}, \bibinfo {author} {\bibfnamefont {S.}~\bibnamefont {McArdle}}, \bibinfo {author} {\bibfnamefont {G.}~\bibnamefont {McCabe}}, \bibinfo {author} {\bibfnamefont {G.}~\bibnamefont {Merton}}, \bibinfo {author} {\bibfnamefont {C.}~\bibnamefont {Miles}}, \bibinfo {author} {\bibfnamefont {A.}~\bibnamefont {Milsted}}, \bibinfo {author} {\bibfnamefont {A.}~\bibnamefont {Mishra}}, \bibinfo {author} {\bibfnamefont {L.}~\bibnamefont {Moncelsi}}, \bibinfo {author} {\bibfnamefont {M.}~\bibnamefont {Naghiloo}}, \bibinfo {author} {\bibfnamefont {K.}~\bibnamefont {Noh}}, \bibinfo {author} {\bibfnamefont {E.}~\bibnamefont {Oblepias}}, \bibinfo {author} {\bibfnamefont {G.}~\bibnamefont {Ortuno}}, \bibinfo {author} {\bibfnamefont {J.~C.}\ \bibnamefont {Owens}}, \bibinfo {author} {\bibfnamefont {J.}~\bibnamefont {Pagdilao}}, \bibinfo {author} {\bibfnamefont {A.}~\bibnamefont {Panduro}}, \bibinfo {author} {\bibfnamefont {J.-P.}\ \bibnamefont {Paquette}}, \bibinfo {author} {\bibfnamefont {R.~N.}\
  \bibnamefont {Patel}}, \bibinfo {author} {\bibfnamefont {G.}~\bibnamefont {Peairs}}, \bibinfo {author} {\bibfnamefont {D.~J.}\ \bibnamefont {Perello}}, \bibinfo {author} {\bibfnamefont {E.~C.}\ \bibnamefont {Peterson}}, \bibinfo {author} {\bibfnamefont {S.}~\bibnamefont {Ponte}}, \bibinfo {author} {\bibfnamefont {H.}~\bibnamefont {Putterman}}, \bibinfo {author} {\bibfnamefont {G.}~\bibnamefont {Refael}}, \bibinfo {author} {\bibfnamefont {P.}~\bibnamefont {Reinhold}}, \bibinfo {author} {\bibfnamefont {R.}~\bibnamefont {Resnick}}, \bibinfo {author} {\bibfnamefont {O.~A.}\ \bibnamefont {Reyna}}, \bibinfo {author} {\bibfnamefont {R.}~\bibnamefont {Rodriguez}}, \bibinfo {author} {\bibfnamefont {J.}~\bibnamefont {Rose}}, \bibinfo {author} {\bibfnamefont {A.~H.}\ \bibnamefont {Rubin}}, \bibinfo {author} {\bibfnamefont {M.}~\bibnamefont {Runyan}}, \bibinfo {author} {\bibfnamefont {C.~A.}\ \bibnamefont {Ryan}}, \bibinfo {author} {\bibfnamefont {A.}~\bibnamefont {Sahmoud}}, \bibinfo {author} {\bibfnamefont
  {T.}~\bibnamefont {Scaffidi}}, \bibinfo {author} {\bibfnamefont {B.}~\bibnamefont {Shah}}, \bibinfo {author} {\bibfnamefont {S.}~\bibnamefont {Siavoshi}}, \bibinfo {author} {\bibfnamefont {P.}~\bibnamefont {Sivarajah}}, \bibinfo {author} {\bibfnamefont {T.}~\bibnamefont {Skogland}}, \bibinfo {author} {\bibfnamefont {C.-J.}\ \bibnamefont {Su}}, \bibinfo {author} {\bibfnamefont {L.~J.}\ \bibnamefont {Swenson}}, \bibinfo {author} {\bibfnamefont {J.}~\bibnamefont {Sylvia}}, \bibinfo {author} {\bibfnamefont {S.~M.}\ \bibnamefont {Teo}}, \bibinfo {author} {\bibfnamefont {A.}~\bibnamefont {Tomada}}, \bibinfo {author} {\bibfnamefont {G.}~\bibnamefont {Torlai}}, \bibinfo {author} {\bibfnamefont {M.}~\bibnamefont {Wistrom}}, \bibinfo {author} {\bibfnamefont {K.}~\bibnamefont {Zhang}}, \bibinfo {author} {\bibfnamefont {I.}~\bibnamefont {Zuk}}, \bibinfo {author} {\bibfnamefont {A.~A.}\ \bibnamefont {Clerk}}, \bibinfo {author} {\bibfnamefont {F.~G. S.~L.}\ \bibnamefont {Brand{\~{a}}o}}, \bibinfo {author} {\bibfnamefont
  {A.}~\bibnamefont {Retzker}},\ and\ \bibinfo {author} {\bibfnamefont {O.}~\bibnamefont {Painter}},\ }\bibfield  {title} {\bibinfo {title} {{Demonstrating a Long-Coherence Dual-Rail Erasure Qubit Using Tunable Transmons}},\ }\href {https://doi.org/10.1103/PhysRevX.14.011051} {\bibfield  {journal} {\bibinfo  {journal} {Phys. Rev. X}\ }\textbf {\bibinfo {volume} {14}},\ \bibinfo {pages} {011051} (\bibinfo {year} {2024})}\BibitemShut {NoStop}%
\bibitem [{\citenamefont {Masluk}\ \emph {et~al.}(2012)\citenamefont {Masluk}, \citenamefont {Pop}, \citenamefont {Kamal}, \citenamefont {Minev},\ and\ \citenamefont {Devoret}}]{Masluk2012}%
  \BibitemOpen
  \bibfield  {author} {\bibinfo {author} {\bibfnamefont {N.~A.}\ \bibnamefont {Masluk}}, \bibinfo {author} {\bibfnamefont {I.~M.}\ \bibnamefont {Pop}}, \bibinfo {author} {\bibfnamefont {A.}~\bibnamefont {Kamal}}, \bibinfo {author} {\bibfnamefont {Z.~K.}\ \bibnamefont {Minev}},\ and\ \bibinfo {author} {\bibfnamefont {M.~H.}\ \bibnamefont {Devoret}},\ }\bibfield  {title} {\bibinfo {title} {Microwave characterization of josephson junction arrays: Implementing a low loss superinductance},\ }\href {https://doi.org/10.1103/PhysRevLett.109.137002} {\bibfield  {journal} {\bibinfo  {journal} {Phys. Rev. Lett.}\ }\textbf {\bibinfo {volume} {109}},\ \bibinfo {pages} {137002} (\bibinfo {year} {2012})}\BibitemShut {NoStop}%
\bibitem [{\citenamefont {Manucharyan}\ \emph {et~al.}(2009)\citenamefont {Manucharyan}, \citenamefont {Koch}, \citenamefont {Glazman},\ and\ \citenamefont {Devoret}}]{Manucharyan2009}%
  \BibitemOpen
  \bibfield  {author} {\bibinfo {author} {\bibfnamefont {V.~E.}\ \bibnamefont {Manucharyan}}, \bibinfo {author} {\bibfnamefont {J.}~\bibnamefont {Koch}}, \bibinfo {author} {\bibfnamefont {L.~I.}\ \bibnamefont {Glazman}},\ and\ \bibinfo {author} {\bibfnamefont {M.~H.}\ \bibnamefont {Devoret}},\ }\bibfield  {title} {\bibinfo {title} {Fluxonium: Single cooper-pair circuit free of charge offsets},\ }\href {https://doi.org/10.1126/science.1175552} {\bibfield  {journal} {\bibinfo  {journal} {Science}\ }\textbf {\bibinfo {volume} {326}},\ \bibinfo {pages} {113} (\bibinfo {year} {2009})}\BibitemShut {NoStop}%
\bibitem [{\citenamefont {Dolan}(1977)}]{Dolan1977}%
  \BibitemOpen
  \bibfield  {author} {\bibinfo {author} {\bibfnamefont {G.~J.}\ \bibnamefont {Dolan}},\ }\bibfield  {title} {\bibinfo {title} {Offset masks for lift‐off photoprocessing},\ }\href {https://doi.org/10.1063/1.89690} {\bibfield  {journal} {\bibinfo  {journal} {Appl. Phys. Lett.}\ }\textbf {\bibinfo {volume} {31}},\ \bibinfo {pages} {337} (\bibinfo {year} {1977})}\BibitemShut {NoStop}%
\bibitem [{\citenamefont {Zhu}\ \emph {et~al.}(2013)\citenamefont {Zhu}, \citenamefont {Ferguson}, \citenamefont {Manucharyan},\ and\ \citenamefont {Koch}}]{Zhu2013}%
  \BibitemOpen
  \bibfield  {author} {\bibinfo {author} {\bibfnamefont {G.}~\bibnamefont {Zhu}}, \bibinfo {author} {\bibfnamefont {D.~G.}\ \bibnamefont {Ferguson}}, \bibinfo {author} {\bibfnamefont {V.~E.}\ \bibnamefont {Manucharyan}},\ and\ \bibinfo {author} {\bibfnamefont {J.}~\bibnamefont {Koch}},\ }\bibfield  {title} {\bibinfo {title} {Circuit qed with fluxonium qubits: Theory of the dispersive regime},\ }\href {https://doi.org/10.1103/PhysRevB.87.024510} {\bibfield  {journal} {\bibinfo  {journal} {Phys. Rev. B}\ }\textbf {\bibinfo {volume} {87}},\ \bibinfo {pages} {024510} (\bibinfo {year} {2013})}\BibitemShut {NoStop}%
\bibitem [{\citenamefont {Smith}\ \emph {et~al.}(2022)\citenamefont {Smith}, \citenamefont {Villiers}, \citenamefont {Marquet}, \citenamefont {Palomo}, \citenamefont {Delbecq}, \citenamefont {Kontos}, \citenamefont {Campagne-Ibarcq}, \citenamefont {Dou\ifmmode~\mbox{\c{c}}\else \c{c}\fi{}ot},\ and\ \citenamefont {Leghtas}}]{Smith2022}%
  \BibitemOpen
  \bibfield  {author} {\bibinfo {author} {\bibfnamefont {W.~C.}\ \bibnamefont {Smith}}, \bibinfo {author} {\bibfnamefont {M.}~\bibnamefont {Villiers}}, \bibinfo {author} {\bibfnamefont {A.}~\bibnamefont {Marquet}}, \bibinfo {author} {\bibfnamefont {J.}~\bibnamefont {Palomo}}, \bibinfo {author} {\bibfnamefont {M.~R.}\ \bibnamefont {Delbecq}}, \bibinfo {author} {\bibfnamefont {T.}~\bibnamefont {Kontos}}, \bibinfo {author} {\bibfnamefont {P.}~\bibnamefont {Campagne-Ibarcq}}, \bibinfo {author} {\bibfnamefont {B.}~\bibnamefont {Dou\ifmmode~\mbox{\c{c}}\else \c{c}\fi{}ot}},\ and\ \bibinfo {author} {\bibfnamefont {Z.}~\bibnamefont {Leghtas}},\ }\bibfield  {title} {\bibinfo {title} {Magnifying quantum phase fluctuations with cooper-pair pairing},\ }\href {https://doi.org/10.1103/PhysRevX.12.021002} {\bibfield  {journal} {\bibinfo  {journal} {Phys. Rev. X}\ }\textbf {\bibinfo {volume} {12}},\ \bibinfo {pages} {021002} (\bibinfo {year} {2022})}\BibitemShut {NoStop}%
\bibitem [{\citenamefont {Wallraff}\ \emph {et~al.}(2007)\citenamefont {Wallraff}, \citenamefont {Schuster}, \citenamefont {Blais}, \citenamefont {Gambetta}, \citenamefont {Schreier}, \citenamefont {Frunzio}, \citenamefont {Devoret}, \citenamefont {Girvin},\ and\ \citenamefont {Schoelkopf}}]{Wallraff2007}%
  \BibitemOpen
  \bibfield  {author} {\bibinfo {author} {\bibfnamefont {A.}~\bibnamefont {Wallraff}}, \bibinfo {author} {\bibfnamefont {D.~I.}\ \bibnamefont {Schuster}}, \bibinfo {author} {\bibfnamefont {A.}~\bibnamefont {Blais}}, \bibinfo {author} {\bibfnamefont {J.~M.}\ \bibnamefont {Gambetta}}, \bibinfo {author} {\bibfnamefont {J.}~\bibnamefont {Schreier}}, \bibinfo {author} {\bibfnamefont {L.}~\bibnamefont {Frunzio}}, \bibinfo {author} {\bibfnamefont {M.~H.}\ \bibnamefont {Devoret}}, \bibinfo {author} {\bibfnamefont {S.~M.}\ \bibnamefont {Girvin}},\ and\ \bibinfo {author} {\bibfnamefont {R.~J.}\ \bibnamefont {Schoelkopf}},\ }\bibfield  {title} {\bibinfo {title} {Sideband transitions and two-tone spectroscopy of a superconducting qubit strongly coupled to an on-chip cavity},\ }\href {https://doi.org/10.1103/PhysRevLett.99.050501} {\bibfield  {journal} {\bibinfo  {journal} {Phys. Rev. Lett.}\ }\textbf {\bibinfo {volume} {99}},\ \bibinfo {pages} {050501} (\bibinfo {year} {2007})}\BibitemShut {NoStop}%
\bibitem [{\citenamefont {Cohen-Tannoudji}\ \emph {et~al.}(1998)\citenamefont {Cohen-Tannoudji}, \citenamefont {Dupont-Roc},\ and\ \citenamefont {Grynberg}}]{cohen1998atom}%
  \BibitemOpen
  \bibfield  {author} {\bibinfo {author} {\bibfnamefont {C.}~\bibnamefont {Cohen-Tannoudji}}, \bibinfo {author} {\bibfnamefont {J.}~\bibnamefont {Dupont-Roc}},\ and\ \bibinfo {author} {\bibfnamefont {G.}~\bibnamefont {Grynberg}},\ }\href {https://onlinelibrary.wiley.com/doi/book/10.1002/9783527617197} {\emph {\bibinfo {title} {Atom-photon interactions: basic processes and applications}}}\ (\bibinfo  {publisher} {John Wiley \& Sons, Ltd},\ \bibinfo {year} {1998})\BibitemShut {NoStop}%
\bibitem [{\citenamefont {Clerk}\ \emph {et~al.}(2010)\citenamefont {Clerk}, \citenamefont {Devoret}, \citenamefont {Girvin}, \citenamefont {Marquardt},\ and\ \citenamefont {Schoelkopf}}]{Clerk2010}%
  \BibitemOpen
  \bibfield  {author} {\bibinfo {author} {\bibfnamefont {A.~A.}\ \bibnamefont {Clerk}}, \bibinfo {author} {\bibfnamefont {M.~H.}\ \bibnamefont {Devoret}}, \bibinfo {author} {\bibfnamefont {S.~M.}\ \bibnamefont {Girvin}}, \bibinfo {author} {\bibfnamefont {F.}~\bibnamefont {Marquardt}},\ and\ \bibinfo {author} {\bibfnamefont {R.~J.}\ \bibnamefont {Schoelkopf}},\ }\bibfield  {title} {\bibinfo {title} {Introduction to quantum noise, measurement, and amplification},\ }\href {https://doi.org/10.1103/RevModPhys.82.1155} {\bibfield  {journal} {\bibinfo  {journal} {Rev. Mod. Phys.}\ }\textbf {\bibinfo {volume} {82}},\ \bibinfo {pages} {1155} (\bibinfo {year} {2010})}\BibitemShut {NoStop}%
\bibitem [{\citenamefont {Dunsworth}\ \emph {et~al.}(2017)\citenamefont {Dunsworth}, \citenamefont {Megrant}, \citenamefont {Quintana}, \citenamefont {Chen}, \citenamefont {Barends}, \citenamefont {Burkett}, \citenamefont {Foxen}, \citenamefont {Chen}, \citenamefont {Chiaro}, \citenamefont {Fowler}, \citenamefont {Graff}, \citenamefont {Jeffrey}, \citenamefont {Kelly}, \citenamefont {Lucero}, \citenamefont {Mutus}, \citenamefont {Neeley}, \citenamefont {Neill}, \citenamefont {Roushan}, \citenamefont {Sank}, \citenamefont {Vainsencher}, \citenamefont {Wenner}, \citenamefont {White},\ and\ \citenamefont {Martinis}}]{Martinis2017}%
  \BibitemOpen
  \bibfield  {author} {\bibinfo {author} {\bibfnamefont {A.}~\bibnamefont {Dunsworth}}, \bibinfo {author} {\bibfnamefont {A.}~\bibnamefont {Megrant}}, \bibinfo {author} {\bibfnamefont {C.}~\bibnamefont {Quintana}}, \bibinfo {author} {\bibfnamefont {Z.}~\bibnamefont {Chen}}, \bibinfo {author} {\bibfnamefont {R.}~\bibnamefont {Barends}}, \bibinfo {author} {\bibfnamefont {B.}~\bibnamefont {Burkett}}, \bibinfo {author} {\bibfnamefont {B.}~\bibnamefont {Foxen}}, \bibinfo {author} {\bibfnamefont {Y.}~\bibnamefont {Chen}}, \bibinfo {author} {\bibfnamefont {B.}~\bibnamefont {Chiaro}}, \bibinfo {author} {\bibfnamefont {A.}~\bibnamefont {Fowler}}, \bibinfo {author} {\bibfnamefont {R.}~\bibnamefont {Graff}}, \bibinfo {author} {\bibfnamefont {E.}~\bibnamefont {Jeffrey}}, \bibinfo {author} {\bibfnamefont {J.}~\bibnamefont {Kelly}}, \bibinfo {author} {\bibfnamefont {E.}~\bibnamefont {Lucero}}, \bibinfo {author} {\bibfnamefont {J.~Y.}\ \bibnamefont {Mutus}}, \bibinfo {author} {\bibfnamefont {M.}~\bibnamefont {Neeley}},
  \bibinfo {author} {\bibfnamefont {C.}~\bibnamefont {Neill}}, \bibinfo {author} {\bibfnamefont {P.}~\bibnamefont {Roushan}}, \bibinfo {author} {\bibfnamefont {D.}~\bibnamefont {Sank}}, \bibinfo {author} {\bibfnamefont {A.}~\bibnamefont {Vainsencher}}, \bibinfo {author} {\bibfnamefont {J.}~\bibnamefont {Wenner}}, \bibinfo {author} {\bibfnamefont {T.~C.}\ \bibnamefont {White}},\ and\ \bibinfo {author} {\bibfnamefont {J.~M.}\ \bibnamefont {Martinis}},\ }\bibfield  {title} {\bibinfo {title} {Characterization and reduction of capacitive loss induced by sub-micron josephson junction fabrication in superconducting qubits},\ }\href {https://doi.org/10.1063/1.4993577} {\bibfield  {journal} {\bibinfo  {journal} {Appl. Phys. Lett.}\ }\textbf {\bibinfo {volume} {111}},\ \bibinfo {pages} {022601} (\bibinfo {year} {2017})}\BibitemShut {NoStop}%
\bibitem [{\citenamefont {Woods}\ \emph {et~al.}(2019)\citenamefont {Woods}, \citenamefont {Calusine}, \citenamefont {Melville}, \citenamefont {Sevi}, \citenamefont {Golden}, \citenamefont {Kim}, \citenamefont {Rosenberg}, \citenamefont {Yoder},\ and\ \citenamefont {Oliver}}]{Oliver2019}%
  \BibitemOpen
  \bibfield  {author} {\bibinfo {author} {\bibfnamefont {W.}~\bibnamefont {Woods}}, \bibinfo {author} {\bibfnamefont {G.}~\bibnamefont {Calusine}}, \bibinfo {author} {\bibfnamefont {A.}~\bibnamefont {Melville}}, \bibinfo {author} {\bibfnamefont {A.}~\bibnamefont {Sevi}}, \bibinfo {author} {\bibfnamefont {E.}~\bibnamefont {Golden}}, \bibinfo {author} {\bibfnamefont {D.}~\bibnamefont {Kim}}, \bibinfo {author} {\bibfnamefont {D.}~\bibnamefont {Rosenberg}}, \bibinfo {author} {\bibfnamefont {J.}~\bibnamefont {Yoder}},\ and\ \bibinfo {author} {\bibfnamefont {W.}~\bibnamefont {Oliver}},\ }\bibfield  {title} {\bibinfo {title} {Determining interface dielectric losses in superconducting coplanar-waveguide resonators},\ }\href {https://doi.org/10.1103/PhysRevApplied.12.014012} {\bibfield  {journal} {\bibinfo  {journal} {Phys. Rev. Appl.}\ }\textbf {\bibinfo {volume} {12}},\ \bibinfo {pages} {014012} (\bibinfo {year} {2019})}\BibitemShut {NoStop}%
\bibitem [{\citenamefont {Gambetta}\ \emph {et~al.}(2017)\citenamefont {Gambetta}, \citenamefont {Murray}, \citenamefont {Fung}, \citenamefont {McClure}, \citenamefont {Dial}, \citenamefont {Shanks}, \citenamefont {Sleight},\ and\ \citenamefont {Steffen}}]{Gambetta2016}%
  \BibitemOpen
  \bibfield  {author} {\bibinfo {author} {\bibfnamefont {J.}~\bibnamefont {Gambetta}}, \bibinfo {author} {\bibfnamefont {C.}~\bibnamefont {Murray}}, \bibinfo {author} {\bibfnamefont {Y.}~\bibnamefont {Fung}}, \bibinfo {author} {\bibfnamefont {D.}~\bibnamefont {McClure}}, \bibinfo {author} {\bibfnamefont {O.}~\bibnamefont {Dial}}, \bibinfo {author} {\bibfnamefont {W.}~\bibnamefont {Shanks}}, \bibinfo {author} {\bibfnamefont {J.}~\bibnamefont {Sleight}},\ and\ \bibinfo {author} {\bibfnamefont {M.}~\bibnamefont {Steffen}},\ }\bibfield  {title} {\bibinfo {title} {Investigating surface loss effects in superconducting transmon qubits},\ }\href {https://doi.org/10.1109/TASC.2016.2629670} {\bibfield  {journal} {\bibinfo  {journal} {IEEE Trans. Appl. Supercond.}\ }\textbf {\bibinfo {volume} {27}},\ \bibinfo {pages} {1} (\bibinfo {year} {2017})}\BibitemShut {NoStop}%
\bibitem [{\citenamefont {Nguyen}\ \emph {et~al.}(2019)\citenamefont {Nguyen}, \citenamefont {Lin}, \citenamefont {Somoroff}, \citenamefont {Mencia}, \citenamefont {Grabon},\ and\ \citenamefont {Manucharyan}}]{nguyenHighCoherenceFluxoniumQubit2019}%
  \BibitemOpen
  \bibfield  {author} {\bibinfo {author} {\bibfnamefont {L.~B.}\ \bibnamefont {Nguyen}}, \bibinfo {author} {\bibfnamefont {Y.-H.}\ \bibnamefont {Lin}}, \bibinfo {author} {\bibfnamefont {A.}~\bibnamefont {Somoroff}}, \bibinfo {author} {\bibfnamefont {R.}~\bibnamefont {Mencia}}, \bibinfo {author} {\bibfnamefont {N.}~\bibnamefont {Grabon}},\ and\ \bibinfo {author} {\bibfnamefont {V.~E.}\ \bibnamefont {Manucharyan}},\ }\bibfield  {title} {\bibinfo {title} {High-{{Coherence Fluxonium Qubit}}},\ }\href {https://doi.org/10.1103/PhysRevX.9.041041} {\bibfield  {journal} {\bibinfo  {journal} {Phys. Rev. X}\ }\textbf {\bibinfo {volume} {9}},\ \bibinfo {pages} {041041} (\bibinfo {year} {2019})}\BibitemShut {NoStop}%
\bibitem [{\citenamefont {Vool}\ \emph {et~al.}(2014)\citenamefont {Vool}, \citenamefont {Pop}, \citenamefont {Sliwa}, \citenamefont {Abdo}, \citenamefont {Wang}, \citenamefont {Brecht}, \citenamefont {Gao}, \citenamefont {Shankar}, \citenamefont {Hatridge}, \citenamefont {Catelani}, \citenamefont {Mirrahimi}, \citenamefont {Frunzio}, \citenamefont {Schoelkopf}, \citenamefont {Glazman},\ and\ \citenamefont {Devoret}}]{Vool2014}%
  \BibitemOpen
  \bibfield  {author} {\bibinfo {author} {\bibfnamefont {U.}~\bibnamefont {Vool}}, \bibinfo {author} {\bibfnamefont {I.~M.}\ \bibnamefont {Pop}}, \bibinfo {author} {\bibfnamefont {K.}~\bibnamefont {Sliwa}}, \bibinfo {author} {\bibfnamefont {B.}~\bibnamefont {Abdo}}, \bibinfo {author} {\bibfnamefont {C.}~\bibnamefont {Wang}}, \bibinfo {author} {\bibfnamefont {T.}~\bibnamefont {Brecht}}, \bibinfo {author} {\bibfnamefont {Y.~Y.}\ \bibnamefont {Gao}}, \bibinfo {author} {\bibfnamefont {S.}~\bibnamefont {Shankar}}, \bibinfo {author} {\bibfnamefont {M.}~\bibnamefont {Hatridge}}, \bibinfo {author} {\bibfnamefont {G.}~\bibnamefont {Catelani}}, \bibinfo {author} {\bibfnamefont {M.}~\bibnamefont {Mirrahimi}}, \bibinfo {author} {\bibfnamefont {L.}~\bibnamefont {Frunzio}}, \bibinfo {author} {\bibfnamefont {R.~J.}\ \bibnamefont {Schoelkopf}}, \bibinfo {author} {\bibfnamefont {L.~I.}\ \bibnamefont {Glazman}},\ and\ \bibinfo {author} {\bibfnamefont {M.~H.}\ \bibnamefont {Devoret}},\ }\bibfield  {title} {\bibinfo {title}
  {Non-poissonian quantum jumps of a fluxonium qubit due to quasiparticle excitations},\ }\href {https://doi.org/10.1103/PhysRevLett.113.247001} {\bibfield  {journal} {\bibinfo  {journal} {Phys. Rev. Lett.}\ }\textbf {\bibinfo {volume} {113}},\ \bibinfo {pages} {247001} (\bibinfo {year} {2014})}\BibitemShut {NoStop}%
\bibitem [{\citenamefont {Pop}\ \emph {et~al.}(2014)\citenamefont {Pop}, \citenamefont {Geerlings}, \citenamefont {Catelani}, \citenamefont {Schoelkopf}, \citenamefont {Glazman},\ and\ \citenamefont {Devoret}}]{Pop2014}%
  \BibitemOpen
  \bibfield  {author} {\bibinfo {author} {\bibfnamefont {I.~M.}\ \bibnamefont {Pop}}, \bibinfo {author} {\bibfnamefont {K.}~\bibnamefont {Geerlings}}, \bibinfo {author} {\bibfnamefont {G.}~\bibnamefont {Catelani}}, \bibinfo {author} {\bibfnamefont {R.~J.}\ \bibnamefont {Schoelkopf}}, \bibinfo {author} {\bibfnamefont {L.~I.}\ \bibnamefont {Glazman}},\ and\ \bibinfo {author} {\bibfnamefont {M.~H.}\ \bibnamefont {Devoret}},\ }\bibfield  {title} {\bibinfo {title} {Coherent suppression of electromagnetic dissipation due to superconducting quasiparticles},\ }\href {https://doi.org/10.1038/nature13017} {\bibfield  {journal} {\bibinfo  {journal} {Nature}\ }\textbf {\bibinfo {volume} {508}},\ \bibinfo {pages} {369} (\bibinfo {year} {2014})}\BibitemShut {NoStop}%
\bibitem [{\citenamefont {Catelani}\ \emph {et~al.}(2011)\citenamefont {Catelani}, \citenamefont {Schoelkopf}, \citenamefont {Devoret},\ and\ \citenamefont {Glazman}}]{Catelani2011}%
  \BibitemOpen
  \bibfield  {author} {\bibinfo {author} {\bibfnamefont {G.}~\bibnamefont {Catelani}}, \bibinfo {author} {\bibfnamefont {R.~J.}\ \bibnamefont {Schoelkopf}}, \bibinfo {author} {\bibfnamefont {M.~H.}\ \bibnamefont {Devoret}},\ and\ \bibinfo {author} {\bibfnamefont {L.~I.}\ \bibnamefont {Glazman}},\ }\bibfield  {title} {\bibinfo {title} {Relaxation and frequency shifts induced by quasiparticles in superconducting qubits},\ }\href {https://doi.org/10.1103/PhysRevB.84.064517} {\bibfield  {journal} {\bibinfo  {journal} {Phys. Rev. B}\ }\textbf {\bibinfo {volume} {84}},\ \bibinfo {pages} {064517} (\bibinfo {year} {2011})}\BibitemShut {NoStop}%
\bibitem [{\citenamefont {Serniak}\ \emph {et~al.}(2018)\citenamefont {Serniak}, \citenamefont {Hays}, \citenamefont {de~Lange}, \citenamefont {Diamond}, \citenamefont {Shankar}, \citenamefont {Burkhart}, \citenamefont {Frunzio}, \citenamefont {Houzet},\ and\ \citenamefont {Devoret}}]{Serniak2018}%
  \BibitemOpen
  \bibfield  {author} {\bibinfo {author} {\bibfnamefont {K.}~\bibnamefont {Serniak}}, \bibinfo {author} {\bibfnamefont {M.}~\bibnamefont {Hays}}, \bibinfo {author} {\bibfnamefont {G.}~\bibnamefont {de~Lange}}, \bibinfo {author} {\bibfnamefont {S.}~\bibnamefont {Diamond}}, \bibinfo {author} {\bibfnamefont {S.}~\bibnamefont {Shankar}}, \bibinfo {author} {\bibfnamefont {L.~D.}\ \bibnamefont {Burkhart}}, \bibinfo {author} {\bibfnamefont {L.}~\bibnamefont {Frunzio}}, \bibinfo {author} {\bibfnamefont {M.}~\bibnamefont {Houzet}},\ and\ \bibinfo {author} {\bibfnamefont {M.~H.}\ \bibnamefont {Devoret}},\ }\bibfield  {title} {\bibinfo {title} {Hot nonequilibrium quasiparticles in transmon qubits},\ }\href {https://doi.org/10.1103/PhysRevLett.121.157701} {\bibfield  {journal} {\bibinfo  {journal} {Phys. Rev. Lett.}\ }\textbf {\bibinfo {volume} {121}},\ \bibinfo {pages} {157701} (\bibinfo {year} {2018})}\BibitemShut {NoStop}%
\bibitem [{\citenamefont {Houck}\ \emph {et~al.}(2008)\citenamefont {Houck}, \citenamefont {Schreier}, \citenamefont {Johnson}, \citenamefont {Chow}, \citenamefont {Koch}, \citenamefont {Gambetta}, \citenamefont {Schuster}, \citenamefont {Frunzio}, \citenamefont {Devoret}, \citenamefont {Girvin},\ and\ \citenamefont {Schoelkopf}}]{Houck2008}%
  \BibitemOpen
  \bibfield  {author} {\bibinfo {author} {\bibfnamefont {A.~A.}\ \bibnamefont {Houck}}, \bibinfo {author} {\bibfnamefont {J.~A.}\ \bibnamefont {Schreier}}, \bibinfo {author} {\bibfnamefont {B.~R.}\ \bibnamefont {Johnson}}, \bibinfo {author} {\bibfnamefont {J.~M.}\ \bibnamefont {Chow}}, \bibinfo {author} {\bibfnamefont {J.}~\bibnamefont {Koch}}, \bibinfo {author} {\bibfnamefont {J.~M.}\ \bibnamefont {Gambetta}}, \bibinfo {author} {\bibfnamefont {D.~I.}\ \bibnamefont {Schuster}}, \bibinfo {author} {\bibfnamefont {L.}~\bibnamefont {Frunzio}}, \bibinfo {author} {\bibfnamefont {M.~H.}\ \bibnamefont {Devoret}}, \bibinfo {author} {\bibfnamefont {S.~M.}\ \bibnamefont {Girvin}},\ and\ \bibinfo {author} {\bibfnamefont {R.~J.}\ \bibnamefont {Schoelkopf}},\ }\bibfield  {title} {\bibinfo {title} {Controlling the spontaneous emission of a superconducting transmon qubit},\ }\href {https://doi.org/10.1103/PhysRevLett.101.080502} {\bibfield  {journal} {\bibinfo  {journal} {Phys. Rev. Lett.}\ }\textbf {\bibinfo {volume}
  {101}},\ \bibinfo {pages} {080502} (\bibinfo {year} {2008})}\BibitemShut {NoStop}%
\bibitem [{\citenamefont {Kittel}(1976)}]{kittel}%
  \BibitemOpen
  \bibfield  {author} {\bibinfo {author} {\bibfnamefont {C.}~\bibnamefont {Kittel}},\ }\href@noop {} {\emph {\bibinfo {title} {Introduction to Solid State Physics, 5th Edition}}}\ (\bibinfo  {publisher} {John Wiley \& Sons, Ltd},\ \bibinfo {year} {1976})\BibitemShut {NoStop}%
\bibitem [{\citenamefont {Koch}\ \emph {et~al.}(2007)\citenamefont {Koch}, \citenamefont {DiVincenzo},\ and\ \citenamefont {Clarke}}]{KochR2007}%
  \BibitemOpen
  \bibfield  {author} {\bibinfo {author} {\bibfnamefont {R.~H.}\ \bibnamefont {Koch}}, \bibinfo {author} {\bibfnamefont {D.~P.}\ \bibnamefont {DiVincenzo}},\ and\ \bibinfo {author} {\bibfnamefont {J.}~\bibnamefont {Clarke}},\ }\bibfield  {title} {\bibinfo {title} {Model for $1/f$ flux noise in squids and qubits},\ }\href {https://doi.org/10.1103/PhysRevLett.98.267003} {\bibfield  {journal} {\bibinfo  {journal} {Phys. Rev. Lett.}\ }\textbf {\bibinfo {volume} {98}},\ \bibinfo {pages} {267003} (\bibinfo {year} {2007})}\BibitemShut {NoStop}%
\bibitem [{\citenamefont {Bialczak}\ \emph {et~al.}(2007)\citenamefont {Bialczak}, \citenamefont {McDermott}, \citenamefont {Ansmann}, \citenamefont {Hofheinz}, \citenamefont {Katz}, \citenamefont {Lucero}, \citenamefont {Neeley}, \citenamefont {O'Connell}, \citenamefont {Wang}, \citenamefont {Cleland},\ and\ \citenamefont {Martinis}}]{Bialczak2007}%
  \BibitemOpen
  \bibfield  {author} {\bibinfo {author} {\bibfnamefont {R.~C.}\ \bibnamefont {Bialczak}}, \bibinfo {author} {\bibfnamefont {R.}~\bibnamefont {McDermott}}, \bibinfo {author} {\bibfnamefont {M.}~\bibnamefont {Ansmann}}, \bibinfo {author} {\bibfnamefont {M.}~\bibnamefont {Hofheinz}}, \bibinfo {author} {\bibfnamefont {N.}~\bibnamefont {Katz}}, \bibinfo {author} {\bibfnamefont {E.}~\bibnamefont {Lucero}}, \bibinfo {author} {\bibfnamefont {M.}~\bibnamefont {Neeley}}, \bibinfo {author} {\bibfnamefont {A.~D.}\ \bibnamefont {O'Connell}}, \bibinfo {author} {\bibfnamefont {H.}~\bibnamefont {Wang}}, \bibinfo {author} {\bibfnamefont {A.~N.}\ \bibnamefont {Cleland}},\ and\ \bibinfo {author} {\bibfnamefont {J.~M.}\ \bibnamefont {Martinis}},\ }\bibfield  {title} {\bibinfo {title} {$1/f$ flux noise in josephson phase qubits},\ }\href {https://doi.org/10.1103/PhysRevLett.99.187006} {\bibfield  {journal} {\bibinfo  {journal} {Phys. Rev. Lett.}\ }\textbf {\bibinfo {volume} {99}},\ \bibinfo {pages} {187006} (\bibinfo {year}
  {2007})}\BibitemShut {NoStop}%
\bibitem [{\citenamefont {Kumar}\ \emph {et~al.}(2016)\citenamefont {Kumar}, \citenamefont {Sendelbach}, \citenamefont {Beck}, \citenamefont {Freeland}, \citenamefont {Wang}, \citenamefont {Wang}, \citenamefont {Yu}, \citenamefont {Wu}, \citenamefont {Pappas},\ and\ \citenamefont {McDermott}}]{Kumar2016}%
  \BibitemOpen
  \bibfield  {author} {\bibinfo {author} {\bibfnamefont {P.}~\bibnamefont {Kumar}}, \bibinfo {author} {\bibfnamefont {S.}~\bibnamefont {Sendelbach}}, \bibinfo {author} {\bibfnamefont {M.~A.}\ \bibnamefont {Beck}}, \bibinfo {author} {\bibfnamefont {J.~W.}\ \bibnamefont {Freeland}}, \bibinfo {author} {\bibfnamefont {Z.}~\bibnamefont {Wang}}, \bibinfo {author} {\bibfnamefont {H.}~\bibnamefont {Wang}}, \bibinfo {author} {\bibfnamefont {C.~C.}\ \bibnamefont {Yu}}, \bibinfo {author} {\bibfnamefont {R.~Q.}\ \bibnamefont {Wu}}, \bibinfo {author} {\bibfnamefont {D.~P.}\ \bibnamefont {Pappas}},\ and\ \bibinfo {author} {\bibfnamefont {R.}~\bibnamefont {McDermott}},\ }\bibfield  {title} {\bibinfo {title} {Origin and reduction of $1/f$ magnetic flux noise in superconducting devices},\ }\href {https://doi.org/10.1103/PhysRevApplied.6.041001} {\bibfield  {journal} {\bibinfo  {journal} {Phys. Rev. Appl.}\ }\textbf {\bibinfo {volume} {6}},\ \bibinfo {pages} {041001} (\bibinfo {year} {2016})}\BibitemShut {NoStop}%
\bibitem [{\citenamefont {Van~Harlingen}\ \emph {et~al.}(2004)\citenamefont {Van~Harlingen}, \citenamefont {Robertson}, \citenamefont {Plourde}, \citenamefont {Reichardt}, \citenamefont {Crane},\ and\ \citenamefont {Clarke}}]{vanharlingen2004}%
  \BibitemOpen
  \bibfield  {author} {\bibinfo {author} {\bibfnamefont {D.~J.}\ \bibnamefont {Van~Harlingen}}, \bibinfo {author} {\bibfnamefont {T.~L.}\ \bibnamefont {Robertson}}, \bibinfo {author} {\bibfnamefont {B.~L.~T.}\ \bibnamefont {Plourde}}, \bibinfo {author} {\bibfnamefont {P.~A.}\ \bibnamefont {Reichardt}}, \bibinfo {author} {\bibfnamefont {T.~A.}\ \bibnamefont {Crane}},\ and\ \bibinfo {author} {\bibfnamefont {J.}~\bibnamefont {Clarke}},\ }\bibfield  {title} {\bibinfo {title} {Decoherence in josephson-junction qubits due to critical-current fluctuations},\ }\href {https://doi.org/10.1103/PhysRevB.70.064517} {\bibfield  {journal} {\bibinfo  {journal} {Phys. Rev. B}\ }\textbf {\bibinfo {volume} {70}},\ \bibinfo {pages} {064517} (\bibinfo {year} {2004})}\BibitemShut {NoStop}%
\bibitem [{\citenamefont {Gambetta}\ \emph {et~al.}(2007)\citenamefont {Gambetta}, \citenamefont {Braff}, \citenamefont {Wallraff}, \citenamefont {Girvin},\ and\ \citenamefont {Schoelkopf}}]{gambettaProtocolsOptimalReadout2007}%
  \BibitemOpen
  \bibfield  {author} {\bibinfo {author} {\bibfnamefont {J.}~\bibnamefont {Gambetta}}, \bibinfo {author} {\bibfnamefont {W.~A.}\ \bibnamefont {Braff}}, \bibinfo {author} {\bibfnamefont {A.}~\bibnamefont {Wallraff}}, \bibinfo {author} {\bibfnamefont {S.~M.}\ \bibnamefont {Girvin}},\ and\ \bibinfo {author} {\bibfnamefont {R.~J.}\ \bibnamefont {Schoelkopf}},\ }\bibfield  {title} {\bibinfo {title} {Protocols for optimal readout of qubits using a continuous quantum nondemolition measurement},\ }\href {https://doi.org/10/cgw553} {\bibfield  {journal} {\bibinfo  {journal} {Phys. Rev. A}\ }\textbf {\bibinfo {volume} {76}},\ \bibinfo {pages} {012325} (\bibinfo {year} {2007})}\BibitemShut {NoStop}%
\bibitem [{\citenamefont {Blais}\ \emph {et~al.}(2021)\citenamefont {Blais}, \citenamefont {Grimsmo}, \citenamefont {Girvin},\ and\ \citenamefont {Wallraff}}]{blaisCircuitQuantumElectrodynamics2021}%
  \BibitemOpen
  \bibfield  {author} {\bibinfo {author} {\bibfnamefont {A.}~\bibnamefont {Blais}}, \bibinfo {author} {\bibfnamefont {A.~L.}\ \bibnamefont {Grimsmo}}, \bibinfo {author} {\bibfnamefont {S.~M.}\ \bibnamefont {Girvin}},\ and\ \bibinfo {author} {\bibfnamefont {A.}~\bibnamefont {Wallraff}},\ }\bibfield  {title} {\bibinfo {title} {Circuit quantum electrodynamics},\ }\href {https://doi.org/10.1103/RevModPhys.93.025005} {\bibfield  {journal} {\bibinfo  {journal} {Rev. Mod. Phys.}\ }\textbf {\bibinfo {volume} {93}},\ \bibinfo {pages} {025005} (\bibinfo {year} {2021})}\BibitemShut {NoStop}%
\bibitem [{\citenamefont {Ding}\ \emph {et~al.}(2024)\citenamefont {Ding}, \citenamefont {Di~Federico}, \citenamefont {Hatridge}, \citenamefont {Houck}, \citenamefont {Leger}, \citenamefont {Martinez}, \citenamefont {Miao}, \citenamefont {I}, \citenamefont {Stefanazzi}, \citenamefont {Stoughton}, \citenamefont {Sussman}, \citenamefont {Treptow}, \citenamefont {Uemura}, \citenamefont {Wilcer}, \citenamefont {Zhang}, \citenamefont {Zhou},\ and\ \citenamefont {Cancelo}}]{Ding2024}%
  \BibitemOpen
  \bibfield  {author} {\bibinfo {author} {\bibfnamefont {C.}~\bibnamefont {Ding}}, \bibinfo {author} {\bibfnamefont {M.}~\bibnamefont {Di~Federico}}, \bibinfo {author} {\bibfnamefont {M.}~\bibnamefont {Hatridge}}, \bibinfo {author} {\bibfnamefont {A.}~\bibnamefont {Houck}}, \bibinfo {author} {\bibfnamefont {S.}~\bibnamefont {Leger}}, \bibinfo {author} {\bibfnamefont {J.}~\bibnamefont {Martinez}}, \bibinfo {author} {\bibfnamefont {C.}~\bibnamefont {Miao}}, \bibinfo {author} {\bibfnamefont {D.~S.}\ \bibnamefont {I}}, \bibinfo {author} {\bibfnamefont {L.}~\bibnamefont {Stefanazzi}}, \bibinfo {author} {\bibfnamefont {C.}~\bibnamefont {Stoughton}}, \bibinfo {author} {\bibfnamefont {S.}~\bibnamefont {Sussman}}, \bibinfo {author} {\bibfnamefont {K.}~\bibnamefont {Treptow}}, \bibinfo {author} {\bibfnamefont {S.}~\bibnamefont {Uemura}}, \bibinfo {author} {\bibfnamefont {N.}~\bibnamefont {Wilcer}}, \bibinfo {author} {\bibfnamefont {H.}~\bibnamefont {Zhang}}, \bibinfo {author} {\bibfnamefont {C.}~\bibnamefont {Zhou}},\
  and\ \bibinfo {author} {\bibfnamefont {G.}~\bibnamefont {Cancelo}},\ }\bibfield  {title} {\bibinfo {title} {Experimental advances with the qick (quantum instrumentation control kit) for superconducting quantum hardware},\ }\href {https://doi.org/10.1103/PhysRevResearch.6.013305} {\bibfield  {journal} {\bibinfo  {journal} {Phys. Rev. Res.}\ }\textbf {\bibinfo {volume} {6}},\ \bibinfo {pages} {013305} (\bibinfo {year} {2024})}\BibitemShut {NoStop}%
\bibitem [{\citenamefont {Stefanazzi}\ \emph {et~al.}(2022)\citenamefont {Stefanazzi}, \citenamefont {Treptow}, \citenamefont {Wilcer}, \citenamefont {Stoughton}, \citenamefont {Bradford}, \citenamefont {Uemura}, \citenamefont {Zorzetti}, \citenamefont {Montella}, \citenamefont {Cancelo}, \citenamefont {Sussman}, \citenamefont {Houck}, \citenamefont {Saxena}, \citenamefont {Arnaldi}, \citenamefont {Agrawal}, \citenamefont {Zhang}, \citenamefont {Ding},\ and\ \citenamefont {Schuster}}]{stefanazzi2022qick}%
  \BibitemOpen
  \bibfield  {author} {\bibinfo {author} {\bibfnamefont {L.}~\bibnamefont {Stefanazzi}}, \bibinfo {author} {\bibfnamefont {K.}~\bibnamefont {Treptow}}, \bibinfo {author} {\bibfnamefont {N.}~\bibnamefont {Wilcer}}, \bibinfo {author} {\bibfnamefont {C.}~\bibnamefont {Stoughton}}, \bibinfo {author} {\bibfnamefont {C.}~\bibnamefont {Bradford}}, \bibinfo {author} {\bibfnamefont {S.}~\bibnamefont {Uemura}}, \bibinfo {author} {\bibfnamefont {S.}~\bibnamefont {Zorzetti}}, \bibinfo {author} {\bibfnamefont {S.}~\bibnamefont {Montella}}, \bibinfo {author} {\bibfnamefont {G.}~\bibnamefont {Cancelo}}, \bibinfo {author} {\bibfnamefont {S.}~\bibnamefont {Sussman}}, \bibinfo {author} {\bibfnamefont {A.}~\bibnamefont {Houck}}, \bibinfo {author} {\bibfnamefont {S.}~\bibnamefont {Saxena}}, \bibinfo {author} {\bibfnamefont {H.}~\bibnamefont {Arnaldi}}, \bibinfo {author} {\bibfnamefont {A.}~\bibnamefont {Agrawal}}, \bibinfo {author} {\bibfnamefont {H.}~\bibnamefont {Zhang}}, \bibinfo {author} {\bibfnamefont {C.}~\bibnamefont
  {Ding}},\ and\ \bibinfo {author} {\bibfnamefont {D.~I.}\ \bibnamefont {Schuster}},\ }\bibfield  {title} {\bibinfo {title} {{The QICK (Quantum Instrumentation Control Kit): Readout and control for qubits and detectors}},\ }\href {https://doi.org/10.1063/5.0076249} {\bibfield  {journal} {\bibinfo  {journal} {Rev. Sci. Instrum.}\ }\textbf {\bibinfo {volume} {93}},\ \bibinfo {pages} {044709} (\bibinfo {year} {2022})}\BibitemShut {NoStop}%
\end{thebibliography}%

\end{document}